\documentclass[aps,pre,notitlepage,superscriptaddress]{revtex4-1}

\usepackage{graphicx}
\usepackage{dcolumn}
\usepackage{bm}
\usepackage[colorlinks= true, linkcolor=blue, citecolor=red, urlcolor=red]{hyperref}
\usepackage{lipsum}
\usepackage{bbold}
\usepackage{mathtools}
\usepackage[normalem]{ulem}
\usepackage{hyperref}
\usepackage{enumitem,comment}

\def\bea{\begin{eqnarray}}
\def\eea{\end{eqnarray}}

\newcommand{\aref}[1]{Appendix \ref{#1}}%

\usepackage{xcolor}

\newcommand{\eref}[1]{Eq.~(\ref{#1})}
\newcommand{\fref}[1]{Fig.~\ref{#1}} 

\begin{document}


\title{Optimal threshold resetting in collective diffusive search}

\author{Arup Biswas}
\email{arupb@imsc.res.in}
\affiliation{The Institute of Mathematical Sciences, CIT Campus, Taramani, Chennai 600113, India}
\affiliation{Homi Bhabha National Institute, Training School Complex, Anushakti Nagar, Mumbai 400094, India}
\affiliation{School of Engineering and Applied Sciences, Harvard University, Cambridge, MA 02138, USA}
\author{Satya N Majumdar}
\email{satyanarayan.majumdar@cnrs.fr}
\affiliation{Laboratoire de Physique Théorique et Modèles Statistiques (LPTMS),
Université de Paris-Sud, Université Paris-Saclay, 91405 Orsay Cedex, France}
\author{Arnab Pal}
\email{Corresponding author: arnabpal@imsc.res.in}
\affiliation{The Institute of Mathematical Sciences, CIT Campus, Taramani, Chennai 600113, India}
\affiliation{Homi Bhabha National Institute, Training School Complex, Anushakti Nagar, Mumbai 400094, India}

\begin{abstract}

Stochastic resetting has attracted significant attention in recent years due to its wide-ranging applications across physics, biology, and search processes. In most existing studies, however, resetting events are governed by an external timer and remain decoupled from the system’s intrinsic dynamics. In a recent Letter by Biswas \textit{et al} \cite{biswas2025target}, we introduced \textit{threshold resetting (TR)} as an alternative, event-driven optimization strategy for target search problems. Under TR, the entire process is reset whenever any searcher reaches a prescribed threshold, thereby coupling the resetting mechanism directly to the internal dynamics. In this work, we study TR-enabled search by $N$ non-interacting \textit{diffusive} searchers in a one-dimensional box $[0,L]$, with the target at the origin and the threshold at $L$. By optimally tuning the scaled threshold distance $u=x_0/L$, the mean first-passage time can be significantly reduced for $N \geq 2$. We identify a critical population size $N_c(u)$ below which TR outperforms reset-free dynamics. Furthermore, we show that, for fixed $u$, the MFPT depends non-monotonically on $N$, attaining a minimum at $N_{\text{opt}}(u)$. We further quantify the achievable speed-up and analyze the operational cost of TR, revealing a nontrivial optimization landscape. These findings highlight threshold resetting as an efficient and realistic optimization mechanism for complex stochastic search processes. 
\end{abstract}

\pacs{Valid PACS appear here}
\maketitle

\section{Introduction}
Over the past decade, stochastic resetting has emerged as a central theme in statistical physics and the study of stochastic processes. In essence, resetting interrupts the natural evolution of a system and restarts it from its initial configuration. This externally imposed intervention, which typically incurs an energetic cost, drives the system out of equilibrium and gives rise to a variety of rich dynamical features. One of its most notable consequences is its ability to significantly enhance first-passage (FP) processes, often leading to faster and more efficient target search. Originally introduced in 2011 \cite{evans_diffusion_2011,evans2011diffusion}, stochastic resetting has since inspired a vast body of work due to its wide-ranging applications ranging from target search problems \cite{evans_stochastic_2020,gupta2022stochastic,kusmierz2014first,reuveni_optimal_2016,pal_first_2017,chechkin2018random,pal2019first,kumar2023universal,pal2019landau,bhat2016stochastic,campos2015phase,pal_search_2020,biswas2024search,nagar2016diffusion,garcia2026target,evans2025stochastic,boyer2017long,pal2025universal,huang2021random}, optimal control \cite{de2023resetting,mori2025optimal}, active motion \cite{sar2023resetting,santra2020run,shee2026steering}, adaptive resetting with machine learning \cite{keidar2025adaptive,meir2025first,bae2025stochastic,munoz2025learning}, kinetic proofreading \cite{ravasio2026evolution,ravasio2026conditioning,biswas2026proofreading},  chemical and biological processes \cite{reuveni2014role,roldan2016stochastic,biswas2023rate,bressloff2020search,rajoria2026broad}, quantum physics \cite{mukherjee2018quantum,yin2023restart,magoni2022emergent}, computer algorithm \cite{loshchilov2016sgdr,asmussen2008asymptotic,church2025accelerating}, even extending to economics \cite{jolakoski2022fate,stojkoski2022income,stojkoski2021geometric} and operation research \cite{bonomo2021mitigating,maurer2001restart,roy2024queues}. Alongside the theoretical development, tabletop experiments \cite{tal2020experimental,besga2020optimal,faisant2021optimal,altshuler2024environmental,vatash2025many,paramanick2024uncovering,goerlich2023experimental,kundu2025emulating} have enabled the confirmation of the theoretical predictions and provided further practical insights and research avenues. 

In the paradigmatic model of resetting \cite{evans_diffusion_2011}, a stochastic process is set to restart at random epochs drawn from an external clock or distribution, independent of the system dynamics. While such protocols have been the subject of extensive research, another pragmatic approach to their implementation is through threshold-crossing events \cite{de2020optimization,de2021optimization}. In here, a resetting event is triggered when the system crosses a predefined threshold. Threshold-crossing events play a pivotal role in a wide range of physical processes, setting a `safety rule' for the system to operate.  A well-known example is the integrate-and-fire neuron model, which captures how neurons respond to stimuli: when the membrane potential crosses a threshold, the neuron fires an action potential before resetting to its resting state \cite{burkitt2006review,bachar2012stochastic,gerstein1964random}. In finance, stop-loss and take-profit strategies rely on predefined thresholds to trigger the sale of assets, minimizing losses or securing profits \cite{kaufman2013trading,shiryaev2007optimal,zhang2001stock,miller1966model}. In physics, fibre bundle models describe systems in which fibres share a common load until a failure threshold is reached, after which ruptures redistribute stress among the remaining fibres \cite{pradhan2010failure,hansen2015fiber}. In software engineering, circuit breakers act as thresholds to prevent systems from repeatedly calling failing services, enhancing stability and resilience \cite{nygard2018release,surendro2021circuit,montesi2018decorator}. Simple stroboscopic threshold mechanisms have also been employed to obtain sustained temporal regularity in chaotic systems with applications to laser modeling \cite{sinha2001using,bhowmick2014targeting}. More recently, it was shown that applying an energy threshold in the Hénon-Heiles Hamiltonian system can accelerate escape dynamics and suppress noise-enhanced stability \cite{cantisan2024energy}. 

While threshold-driven processes have been extensively studied across disciplines, relatively little attention has been paid to how thresholds influence stochastic search dynamics. In \cite{de2020optimization,de2021optimization}, the authors focus on the spatial properties of a single particle under the threshold resetting (TR) strategy. Recently, the spatial properties  such as the steady state, order statistics and counting statistics of $N$ non-interacting diffusing particles under TR have been investigated in \cite{biroli2026first}.  Bressloff considered the desorption of a single particle from a sticky boundary as a threshold resetting event, but results there focused on the optimization with respect to bulk resetting rate \cite{bressloff2025random,bressloff2026renewal}. In \cite{garcia2026target}, the authors consider a distribution of thresholds in a target search, but the primary focus was again on the optimization with respect to the bulk resetting rate. In contrast, in \cite{biswas2025target}, we proposed a FP optimization technique based on TR. In that study, we considered a FP process conducted by a group of searchers in parallel. Furthermore, the process is interrupted when any of the searchers reaches a threshold. Upon hitting the threshold, a system-wide global reset is triggered, after which all the searchers return to the initial position from where the search resumes. Such a collective reset strategy has natural prevalence in a myriad of processes, such as dynamics of fish schools \cite{couzin2005effective,couzin2011uninformed,conradt2005consensus}, swarming robots \cite{brambilla2013swarm,lerman2004review} and clonal raider ants (\textit{Ooceraea biroi}) \cite{chase2025physics}. A similar rationale was analytically studied much earlier in the context of operation research \cite{asmussen2008asymptotic} where a job is subdivided into $N$ parallel components and subject to restart upon failure. {\color{black} We also note that TR can be conceptually viewed as a limiting case of space-dependent resetting \cite{evans2011diffusion} or adaptive resetting \cite{keidar2025adaptive}. Although such a correspondence can be established in principle, it may not be very useful for practical purposes.}  Quite interestingly, such simultaneous resetting introduces strong correlations among the otherwise independent searchers, effectively coupling their dynamics \cite{biroli2023extreme,biroli2023critical,biroli2024exact,boyer2025emerging,de2026dynamically,biroli2024dynamically,biroli2026experimental,sabhapandit2024noninteracting,mesquita2025dynamically,galla2026diffusion,olsen2026information} (see \cite{biroli2026experimental} for an experimental demonstration of this dynamically emergent correlation mechanism and \cite{majumdar2026dynamically} for a perspective on this topic). In \cite{biswas2025target}, a general framework was developed to analyze any arbitrary search processes under TR. Using a simple model of \textit{ballistic} searchers with randomly assigned initial velocities in one dimension, we demonstrated that TR can yield rich and nontrivial optimization behavior in the search time.

The aim of the present study is to investigate the impact of the TR mechanism on a target search process involving 
$N$ non-interacting \textit{diffusive} searchers. Our primary focus is on the fastest first-passage time—the time taken by the earliest among the 
$N$ searchers to reach the target—a quantity of central interest in extreme value statistics \cite{majumdar2020extreme}. In the absence of any resetting or threshold mechanism, it is known that the mean first-passage time (MFPT) for diffusive searchers becomes finite only for $N \geq 3$, and diverges otherwise \cite{lindenberg1980lattice}. This raises a natural question: Can resetting improve the efficiency of the diffusive search process? Recent studies have shown that standard stochastic resetting can indeed reduce the MFPT, with an optimal resetting rate enhancing search performance \cite{biroli2023critical}. In particular, 
 two different search
protocols were studied in \cite{biroli2023critical} -- when each of them resets (a)
independently and (b) simultaneously. In both the cases, it was shown that there exists a critical number of searchers $N_c$ below which
resetting is beneficial. In particular, for the case (a), it was found that $N_c=8$, which was also derived separately in \cite{bhat2016stochastic}. On the other hand, for the case (b), it was shown that $N_c=7$ \cite{biroli2023critical}.
Complementary to that work, the current paper addresses the following questions: How does first-passage optimization manifest in diffusive search when the simultaneous resetting protocol (similar to case (b)) is internally generated, triggered by a threshold rather than externally controlled? In particular, how does the trade-off between the number of diffusive walkers and the threshold distance influence performance? While some preliminary results for diffusive walkers were reported in \cite{biswas2025target}, the primary focus there was on ballistic search. In contrast, here we employ the general formalism developed in \cite{biswas2025target} to provide a detailed analysis of diffusive search and uncover its rich optimization behavior with respect to both the threshold and the number of searchers.

\begin{figure}
    \centering
    \includegraphics[width=8.5cm]{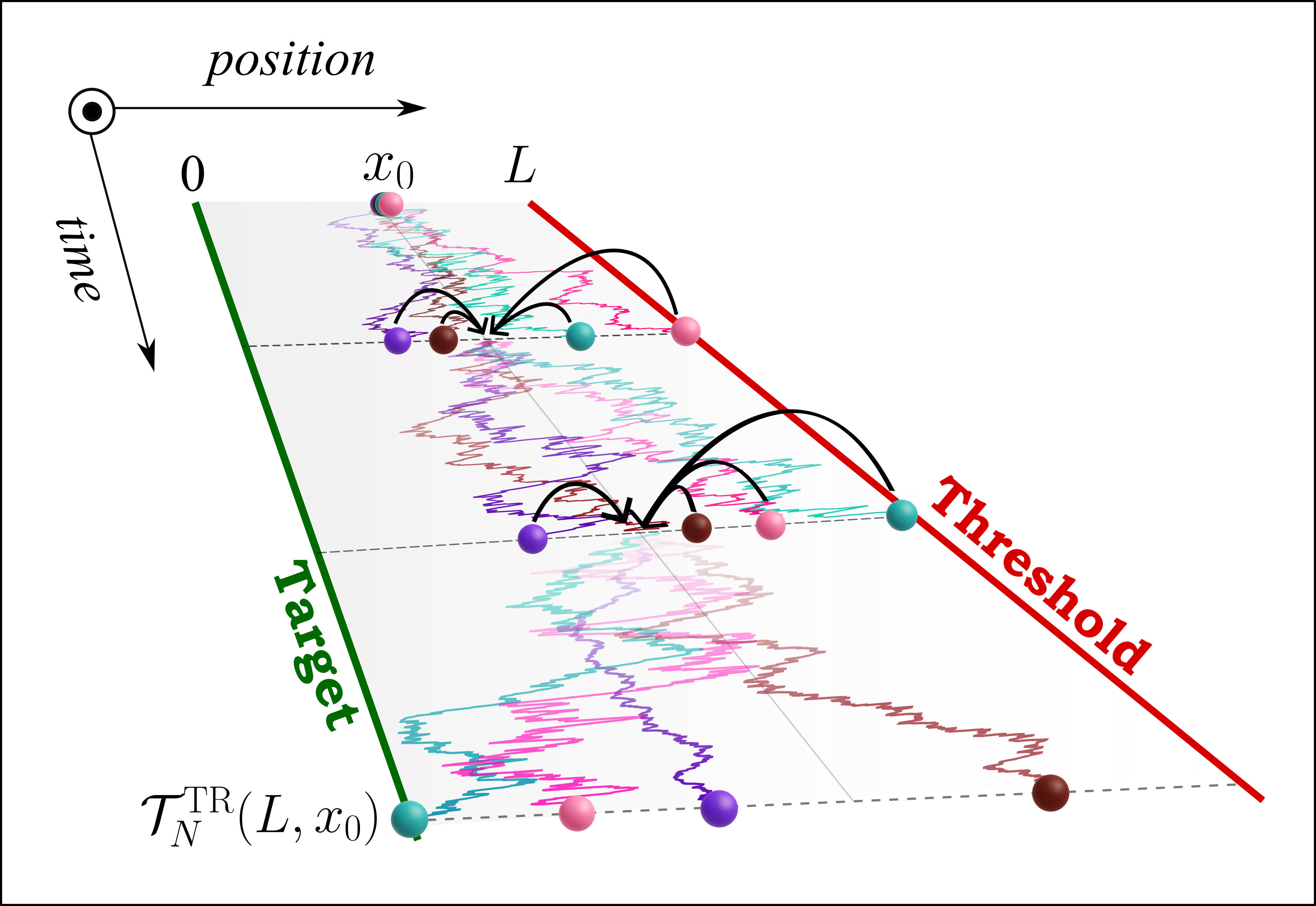}
    \caption{Schematic representation of threshold resetting with $N=4$ non-interacting diffusive searchers. If any one of them reaches the target at $x=0$ we mark the process as complete and note the associated first passage time $ \mathcal{T}_N^{\text{TR}}(L,x_0)$. However, if any of them reaches the threshold at $L$ first, then all of them are simultaneously reset to $x_0$, from where they renew their search. Our aim is to compute the MFPT to the target by these diffusive searchers.
    }
    \label{fig1}
\end{figure}

To address these questions systematically, we consider $N$ diffusive searchers (starting from $x_0$) in one-dimension engaged in locating the target at $x=0$ with the threshold being at $x=L$ (see \fref{fig1}). For a single diffusive searcher, one would expect the MFPT to decrease monotonically with reducing $L$ since the searcher can not escape far away from the target. However, we find MFPT shows non-monotonic behaviour with respect to $L$ whenever the number of searchers is greater than unity. Analogous to tuning the resetting rate, we show the existence of a scaled optimal threshold distance $u_\text{opt}=x_0/L$ that significantly accelerates the search for $N\ge 2$. Moreover, we identify a critical number of searchers $N=N_c(u)$ below which TR consistently yields a lower MFPT than the no-reset scenario. Combined together, we provide a phase diagram in the $u-N$ plane that captures the region where TR expedites the reset-free process. Next, we optimize the MFPT with respect to $N$ that yields an optimal number of searchers $N_{\text{opt}}(u)$ for which the collective search is always useful. \textcolor{black}{We carry out an asymptotic analysis of the MFPT, demonstrating the existence of a boundary layer when the initial position lies in close proximity to the threshold.} We then analyze the speed-up ratios which capture how much faster the search becomes when the TR strategy is employed at its optimal threshold, compared to a solo and resetting-free diffusive process, respectively.  Interestingly, our results demonstrate that an optimally tuned threshold-resetting protocol yields a more efficient collective search than the corresponding optimally tuned Poissonian resetting protocol. Finally, we quantify the cost of TR search, following \cite{biswas2025target}, and demonstrate its rich optimization landscape with respect to the threshold parameter. In particular, we show that the cost function is optimal with respect to $u$ for any number of searchers. In summary, the present work provides a comprehensive analysis of threshold-mediated optimization in diffusive systems, elucidating the interplay between collective dynamics, and threshold to the first-passage efficiency.

The paper is organized as follows; First, we describe the problem set-up in detail in Sec. \ref{setup}. Then we propose two different formalisms to extract the first-passage statistics of the search process under TR in Sec. \ref{formalism-I} and \ref{formalism-II}. Then, in Sec. \ref{eg-diffusion}, we illustrate the formalism presented earlier with the example of $N$ independent diffusive searchers. In particular, we discuss in detail the variation of the MFPT with respect to the threshold distance in Sec. \ref{var-threshold} and the number of searchers in Sec. \ref{var-searchers}. The optimal speed-up due to collective TR is discussed in Sec. \ref{speed-up}. Finally, we define the cost function associated with the target search process with TR and discuss its behaviour for the diffusive search process in Sec. \ref{costs}. We conclude with a brief discussion in Sec. \ref{disc}.

\subsection*{Notations used in the article}
Before proceeding to the main results, we briefly introduce the key notations used throughout the paper for the reader’s convenience. 
\begin{itemize}
    \item $~\mathcal{T}_N^{\text{TR}}(L,x_0) \equiv $ Fastest first passage time of $N$ searchers to reach the target at $x=0$ starting from $x=x_0$ in presence of the threshold at $x=L$ (described in detail in Sec. \ref{setup}). This is the primary observable of interest in our study.
    \item $~T_{N}(L,x_0) \equiv $ Unconditional fastest first passage time of any one of $N$ searchers to reach either of the target or threshold.

    \item $~ q_N(L,x_0,t) \equiv $ Survival probability that none of the $N$ searchers has found either the threshold or the target up to time $t$.

    \item  $\mathcal{Q}_N^{\text{TR}}(L,x_0,t) \equiv $ Survival probability of $N$ searchers under TR. This estimates the probability that none of the $N$ searchers has hit the target up to time $t$.

    \item $~ j_{L,N}(L,x_0,t)~  (\text{or } j_{0,N}(L,x_0,t)) \equiv $ probability current contributed by any of the $N$ searchers reaching the threshold (or target) at time $t$.

    \item $~\epsilon_L (N,L,x_0) ~(\text{or }\epsilon_0 (N,L,x_0)) \equiv $ splitting probability for any of the $N$ searchers to first reach the threshold at the $x=L$ (target at $x=0$) before hitting the target (threshold).

     \item $~Q(x_0,t) \equiv $ Single searcher survival probability of not finding either the target or the threshold up to time $t$.

\end{itemize}

\section{Problem set-up and methodology} 
\label{setup}

We consider a system of $N$ non-interacting stochastic searchers confined to the finite interval $x \in [0, L]$, each initialized at the same position $x_0 \in (0, L)$ (see Fig. \ref{fig1}). The searchers evolve independently according to a general stochastic dynamics, which can encompass diffusion or other random motion models. The left boundary at $x = 0$ is designated as the \textit{target}, and the search is deemed successful—and the process terminates—as soon as \textit{any one} of the $N$ searchers reaches this point. In contrast, the right boundary at $x = L$ serves as a \textit{resetting threshold}. Whenever \textit{any} of the searchers hits this boundary, a global reset is triggered: all $N$ searchers are simultaneously returned to the initial position $x_0$, and the search restarts from this configuration. This mechanism defines the \textit{threshold resetting} (TR) protocol under investigation. Our primary objective is to compute and analyze the statistical properties of the first-passage time, denoted by $\mathcal{T}_N^{\text{TR}}$, to the target under this TR mechanism. To this end, we will adapt two complementary renewal approaches: first, as delineated in \cite{biswas2025target} using survival probability, and second, using the first-passage time random variables directly as formulated in \cite{pal_first_2017}.

\subsection{Formalism I- Renewal approach with survival probability} \label{formalism-I}
We start by denoting the survival probability of the $N$ searcher system under TR by $\mathcal{Q}_N^{\text{TR}}(L,x_0,t)$. This measures the probability that none of the searchers has found the target at $x=0$ up to time $t$, starting from $x_0$. They can, however, hit the resetting boundary at $x=L$ once or multiple times. It is thus also useful to introduce $q_N(L,x_0,t)$ as the survival probability that none of the searchers has hit either of the boundaries at $x=0$ or $x=L$ up to time $t$. With these definitions at hand, one can write the following renewal equation for the survival probability under TR mechanism \cite{biswas2025target}
\begin{align}
   &\mathcal{Q}_N^{\text{TR}}(L,x_0,t)=q_N(L,x_0,t)+\int_{0}^t dt' j_{L,N}(L,x_0,t') \mathcal{Q}_N^{\text{TR}}(L,x_0,t-t'),
   \label{surv-ren}
\end{align}
where $j_{L,N}(L,x_0,t')$ is the probability current contributed by any of the $N$ searchers reaching the threshold $L$ at time $t'$. The physical interpretation of the above equation is as follows: Starting from the position $x_0$, all the $N$ searchers can survive up to time $t$ in two ways. First, the searchers survive without hitting either of the targets, which occurs with the probability $q_N(L,x_0,t)$ -- this is the ``no resetting'' scenario. Secondly, the searchers may hit the threshold at $L$ multiple times without hitting the target up to time $t$. In that case, whenever any one of the $N$ searchers hits the threshold for the very \textit{first time} (say at $t'$), all the searchers are instantaneously reset to the starting position $x_0$, and the process renews. The contribution for any searcher hitting the resetting boundary at $L$ is essentially the flux $j_{L,N}(L,x_0,t')$ followed by the survival probability $\mathcal{Q}_N^{\text{TR}}(L,x_0,t-t')$ for the remaining duration. 
Taking the Laplace transform of the renewal equation \eref{surv-ren} and rearranging, we find
\begin{align}
    \widetilde{\mathcal{Q}}_N^{\text{TR}}(L,x_0,s)= \frac{\widetilde{q}_N(L,x_0,s)}{1-\widetilde{j}_{L,N}(L,x_0,s)}, \label{surv}
\end{align}
where we define the Laplace transformed quantities as, $ \widetilde{\mathcal{Q}}_N^{\text{TR}}(L,x_0,s)=\int_0^\infty dt~ e^{-st}  \mathcal{Q}_N^{\text{TR}}(L,x_0,t)$, $\widetilde{q}_N(L,x_0,s)=\int_0^\infty dt~e^{-s t} q_N(L,x_0,t),$ and $\widetilde{j}_{L,N}(L,x_0,s)= \int_0^\infty dt~e^{-s t} j_{L,N}(L,x_0,t)$.

Importantly, both the survival probability $q_N(L,x_0,t)$ and  the flux $j_{L,N}(L,x_0,t)$ can be expressed in terms of the single searcher observables, assuming they are non-interacting in nature, in the following way
\begin{align}
q_N(L,x_0,t) &= \left[ Q(L,x_0,t) \right]^N,\label{qng}\\
    j_{L,N}(L,x_0,t)&=N j_{L,1}(L,x_0,t) \left[ Q(L,x_0,t) \right]^{N-1},\label{jng}
\end{align}
where $Q(L,x_0,t)$ is the survival probability for a single searcher in the interval $[0,L]$ and $j_{L,1}(L,x_0,t)$ is the single-searcher probability current through the threshold at $x=L$. We note that the
first passage time density $f_{\mathcal{T}_{N}^{\text{TR}}}(L,x_0,t)$ is related to the survival probability through the relation $$f_{\mathcal{T}_{N}^{\text{TR}}}(L,x_0,t)=-\frac{\partial \mathcal{Q}_N^{\text{TR}}(L,x_0,t)}{\partial t},$$ which, in the Laplace space, translates to 
\begin{align}
     \langle e^{-s \mathcal{T}_{N}^{\text{TR}}} \rangle & \equiv \int_0^{\infty} dt ~e^{-st} f_{\mathcal{T}_{N}^{\text{TR}}}(L,x_0,t)  \nonumber\\
    & =1-s\widetilde{\mathcal{Q}}_N^{\text{TR}}(L,x_0,s).
    \label{surv-mgf}
\end{align}
Finally, the $m^{th}$ moment of the first passage time can be obtained from the following relation
\begin{align}
    \left \langle [\mathcal{T}_N^{\text{TR}} (L,x_0)]^m \right\rangle &= (-1)^m \lim_{s\to 0}\frac{\partial^m  }{ \partial s^m}\left[\langle e^{-s \mathcal{T}_{N}^{\text{TR}}} \rangle\right] \nonumber \\
    &=(-1)^{m+1} m\left.\frac{\partial^{m-1} \widetilde{\mathcal{Q}}_N^{\text{TR}}(L,x_0,s)}{\partial s^{m-1}} \right|_{s\to 0}. \label{mmoment}
\end{align}
In what follows, we explicitly derive an exact expression for the mean first passage time (MFPT) starting from \eref{mmoment}.
\vspace{0.5cm}

\subsection*{\textbf{\textit{Mean first passage time}}}
The mean first passage time, following \eref{mmoment}, reads
\begin{align}
    \langle \mathcal{T}_N^{\text{TR}}(L,x_0) \rangle& =   \widetilde{\mathcal{Q}}_N^{\text{TR}}(L,x_0,s\to 0) =\frac{\langle T_{N}(L,x_0) \rangle}{\epsilon_0(N,L,x_0)}, \label{mfpt-1}
\end{align}
where $\langle T_{N}(L,x_0) \rangle$ is the \textit{unconditional} MFPT  for any of the $N$ searchers to reach either the threshold or the target and is given by
\begin{align}
    \langle T_{N}(L,x_0) \rangle=\int_0^\infty dt~ \left[Q(L,x_0,t) \right]^N. \label{unc-mfpt}
\end{align}
On the other hand, $\epsilon_0(N,L,x_0)=1-\epsilon_L(N,L,x_0)=1-\int_0^\infty dt ~j_{L,N}(L,x_0,t)=\int_0^\infty dt ~j_{0,N}(L,x_0,t)$
is simply the splitting probability for any of the $N$ searchers to first reach the target at the origin before hitting the threshold at $L$.
In terms of the single searcher observables, this splitting probability can be written in the following form
\cite{krapivsky2010maximum}
\begin{align}
  \epsilon_0(N,L,x_0)
  &=\int_0^\infty dt~N j_{0,1}(L,x_0,t) [Q(L,x_0,t)]^{N-1}, \label{exit-o}
\end{align}
where $j_{0,1}(L,x_0,t)$ is the single searcher probability current through the target at $x=0$. Finally, combining all the above results we arrive at
\begin{align}
     \langle \mathcal{T}_N^{\text{TR}}(L,x_0) \rangle =\frac{\int_0^\infty dt~ [Q(L,x_0,t)]^N}{\int_0^\infty dt~N j_{0,1}(L,x_0,t) [Q(L,x_0,t)]^{N-1}}, \label{mfpt}
\end{align}
which was derived in \cite{biswas2025target}. The derived expression is neither particular to any underlying dynamics nor does it depend on the precise nature of the resetting boundary. Finally, although the formalism is illustrated using a one-dimensional set-up, the results are general and hold in higher dimensions as well. 

\subsection{Formalism II- Renewal approach with first-passage time variables} \label{formalism-II}
In this section, we provide a complementary approach based on the renewal theory of first-passage time random variables to extract the statistics.  Similar approaches have been taken earlier in other resetting related works \cite{reuveni_optimal_2016,pal_first_2017,pal_search_2020,bressloff2020search} and in gated search problems \cite{bressloff2015escape,bressloff2016diffusion}. We start by denoting $\mathcal{T}_{0,N}$ as the random variable associated with the FPT by any of the $N$ searchers to the target at $x=0$. Similarly, we denote $\mathcal{T}_{L,N}$ as the FPT by any of the $N$ searchers to hit the threshold. With these definitions, one can then write the renewal equation for the random variable associated with the first-passage time of the complete process $\mathcal{T}_{N}^{\text{TR}}$ as
\begin{align}
\begin{array}{lll}
\mathcal{T}_{N}^{\text{TR}}=\left\{ \begin{array}{ll}
\mathcal{T}_{0,N}, ~ & ~\text{if } ~\mathcal{T}_{0,N}<\mathcal{T}_{L,N} \vspace{0.2cm}\\
\mathcal{T}_{L,N}+\mathcal{T}_{N}^{'\text{TR}},~ &~ \text{if }~\mathcal{T}_{L,N} \leq \mathcal{T}_{0,N}  \end{array}\right.\text{ }\end{array},
\label{renewal-1}
\end{align}
where $\mathcal{T}_{N}^{'\text{TR}}$ is an independent and identically distributed (i.i.d.) copy of $\mathcal{T}_{N}^{\text{TR}}$. The above equation can be interpreted in the following way. Starting from an initial condition, if the searchers find the target before hitting the threshold, the process therein ends and thus $\mathcal{T}_{N}^{\text{TR}}=\mathcal{T}_{0,N}$. Otherwise, upon reaching the threshold prior to the target, all the searchers are simultaneously brought back to the resetting coordinate from where the search renews. This is reflected into the second condition of Eq. (\ref{renewal-1}). 
\eref{renewal-1} can be written in a more compact way as
\begin{align}
   & \mathcal{T}_{N}^{\text{TR}}=I(\mathcal{T}_{0,N}<\mathcal{T}_{L,N})\mathcal{T}_{0,N}+ I(\mathcal{T}_{L,N} \le \mathcal{T}_{0,N})[\mathcal{T}_{L,N}+\mathcal{T}_{N}^{'\text{TR}}],
\end{align}
where $I(X<Y)$ denotes an indicator function which takes the value unity only when $X<Y$, otherwise zero. We can now find the moment-generating function of the random variable $\mathcal{T}_{N}^{\text{TR}}$ as
\begin{align}
    \langle e^{-s \mathcal{T}_{N}^{\text{TR}}} \rangle &= \langle I(\mathcal{T}_{0,N}<\mathcal{T}_{L,N}) e^{-s\mathcal{T}_{0,N}}\rangle + \langle I(\mathcal{T}_{L,N}\le \mathcal{T}_{0,N}) e^{-s[\mathcal{T}_{L,N}+\mathcal{T}_{N}^{'\text{TR}}]} \rangle \nonumber\\
    &=\langle I(\mathcal{T}_{0,N}<\mathcal{T}_{L,N}) e^{-s\mathcal{T}_{0,N}}\rangle  + \langle I(\mathcal{T}_{L,N}\le \mathcal{T}_{0,N}) e^{-s\mathcal{T}_{L,N}} \rangle \langle e^{-s \mathcal{T}_{N}^{'\text{TR}}} \rangle \nonumber\\
    &=\langle I(\mathcal{T}_{0,N}<\mathcal{T}_{L,N}) e^{-s\mathcal{T}_{0,N}}\rangle  + \langle I(\mathcal{T}_{L,N}\le \mathcal{T}_{0,N}) e^{-s\mathcal{T}_{L,N}} \rangle \langle e^{-s \mathcal{T}_{N}^{\text{TR}}} \rangle ,
\end{align}
where we have used the independent and identical property of $\mathcal{T}_{N}^{\text{TR}}$ and $\mathcal{T}_{N}^{'\text{TR}}$. Rearranging, we find
\begin{align}
     \langle e^{-s \mathcal{T}_{N}^{\text{TR}}} \rangle= \frac{\langle I(\mathcal{T}_{0,N}<\mathcal{T}_{L,N}) e^{-s\mathcal{T}_{0,N}}\rangle}{1-\langle I(\mathcal{T}_{L,N}\le \mathcal{T}_{0,N}) e^{-s\mathcal{T}_{L,N}} \rangle}. \label{ren-2}
\end{align}
Our next task is to compute the numerator and denominator written in terms of the indicator functions. Let us compute the numerator first, which is
\begin{align}
    &\langle I(\mathcal{T}_{0,N}<\mathcal{T}_{L,N}) e^{-s\mathcal{T}_{0,N}}\rangle= \text{Pr}(\mathcal{T}_{0,N}<\mathcal{T}_{L,N}) \langle e^{-s \{\mathcal{T}_{0,N}|\mathcal{T}_{0,N}<\mathcal{T}_{L,N}\}} \rangle.
    \label{Indicator-II}
\end{align}
Here $\text{Pr}(\mathcal{T}_{0,N}<\mathcal{T}_{L,N})$ stands for the probability that $\mathcal{T}_{0,N}<\mathcal{T}_{L,N}$, which is essentially the splitting probability $\epsilon_0(N,L,x_0)$ defined in Eq. (\ref{exit-o}).
The quantity $\mathcal{T}_{0,N}^c \equiv \{\mathcal{T}_{0,N}|\mathcal{T}_{0,N}<\mathcal{T}_{L,N}\}$ is the conditional time for any of the $N$ searchers to reach the target first before hitting the threshold at $x=L$ and its normalized density can be written as \cite{redner2001}
\begin{align}
    f_{\mathcal{T}_{0,N}^c}(t)=\frac{j_{0,N}(L,x_0,t)}{\text{Pr}(\mathcal{T}_{0,N}<\mathcal{T}_{L,N})}, \label{toden}
\end{align}
where $j_{0,N}(L,x_0,t)$ is the probability current at the target $x=0$  at time $t$ due to any one of the $N$ searchers. Using the above-mentioned definition in Eq. (\ref{toden}), we arrive at the result (see \aref{formalism-II-Appendix})
\begin{align}
    \langle I(\mathcal{T}_{0,N}<\mathcal{T}_{L,N}) e^{-s\mathcal{T}_{0,N}}\rangle&=\int_0^\infty dt~ e^{-s t} j_{0,N}(L,x_0,t) \nonumber\\
    &=\widetilde{j}_{0,N}(L,x_0,s). \label{jos}
\end{align}
Similarly, following  \aref{formalism-II-Appendix}, we find 
\begin{align}
    \langle I(\mathcal{T}_{L,N}\le \mathcal{T}_{0,N}) e^{-s\mathcal{T}_{L,N}} \rangle=\widetilde{j}_{L,N}(L,x_0,s). \label{jls}
\end{align}
Finally, combining all those results, we have the generating function of the FPT given by
\begin{align}
       \langle e^{-s \mathcal{T}_{N}^{\text{TR}}} \rangle =\frac{\widetilde{j}_{0,N}(L,x_0,s)}{1-\widetilde{j}_{L,N}(L,x_0,s)}. \label{mgf}
\end{align}
In terms of the survival probability, this reads
\begin{align}
    \widetilde{\mathcal{Q}}_N^{\text{TR}}(L,x_0,s)&=\frac{1}{s}\left(1- \langle e^{-s \mathcal{T}_{N}^{\text{TR}}} \rangle\right)
    \nonumber\\
    &=\frac{1}{s}\frac{1-[\widetilde{j}_{0,N}(L,x_0,s)+\widetilde{j}_{L,N}(L,x_0,s)]}{1-\widetilde{j}_{L,N}(L,x_0,s)}, \label{surv-11}
\end{align}
which leads to \eref{surv} due to the following relation between the survival probability and the currents (see \aref{appa} for details)
\begin{align}
    \widetilde{q}_N(L,x_0,s) 
    =\frac{1}{s}\left[1- (\widetilde{j}_{0,N}(L,x_0,s)+\widetilde{j}_{L,N}(L,x_0,s))\right], 
    \label{surv-und}
\end{align}
thus justifying the equivalence between the two formalisms.   In what follows, we consider $N$ non-interacting diffusive searchers in one spatial dimension (1D), and by applying the methodology developed above, we derive explicit results and reveal nontrivial features of the search dynamics under the TR protocol.\\

\section{Optimized Diffusive search} \label{eg-diffusion}
Let us consider the case of $N$ diffusive searchers with the same set-up as shown in \fref{fig1}. To proceed, we will need the single searcher diffusive propagator $G(x,t|x_0)$ in the presence of two absorbing boundaries at $x=0$ and $x=L$. This is well known in the literature and is given by \cite{redner2001}
\begin{align}
    G(x,t|x_0)=\frac{2}{L} \sum _{n=0}^{\infty}  \sin \left(\frac{\pi  n x}{L}\right) \sin \left(\frac{\pi  n x_0}{L}\right) e^{-\frac{n^2 \pi ^2 D t}{L^2}}, \label{single-prop}
\end{align}
where $D$ is the diffusion constant. The single searcher survival probability $Q(L,x_0,t)$, the flux through the origin $ j_{0,1}(L,x_0,t)$ and the threshold $j_{L,1}(L,x_0,t)$ respectively can be obtained using the following relations
\begin{align}
   & Q(L,x_0,t)=\int_0^L G(x,t|x_0)dx, \\
    & j_{0,1}(L,x_0,t)=D\left.\frac{\partial G(x,t|x_0)}{\partial x}\right|_{x=0}, \\
     &j_{L,1}(L,x_0,t)  =-D\left.\frac{\partial G(x,t|x_0)}{\partial x}\right|_{x=L}.
\end{align}

\noindent
Using the exact expression of the propagator for the diffusive searchers, we find
\begin{align}
    &Q(u,t)=\frac{2}{\pi}\sum _{n=1}^{\infty} \left(\frac{1-(-1)^n }{  n} \right)\sin \left(n\pi u\right) e^{-\frac{n^2 \pi ^2 t}{\tau_d}},
    \label{surv-nd}\\
    & j_{0,1}(u,t)=\frac{2 \pi }{\tau_d} \sum _{n=1}^{\infty} n \sin  \left(n\pi u\right) e^{-\frac{n^2 \pi ^2  t}{\tau_d}},\label{jond}\\
  & j_{L,1}(u,t) =-\frac{2 \pi}{\tau_d} \sum _{n=1}^{\infty} n (-1)^n \sin  \left(n\pi u\right) e^{-\frac{n^2 \pi ^2 t}{\tau_d}}, \label{jlnd}
\end{align}
where we have introduced
\begin{align}
    u=\frac{x_0}{L}, ~~~~~ \tau_d=\frac{L^2}{D}.
\end{align}
Evidently, $u$ is a dimensionless variable with $0 \leq u \leq 1$ while
$\tau_d$ is the diffusive time scale.
\begin{figure*}
     \centering
     \includegraphics[width=17cm]{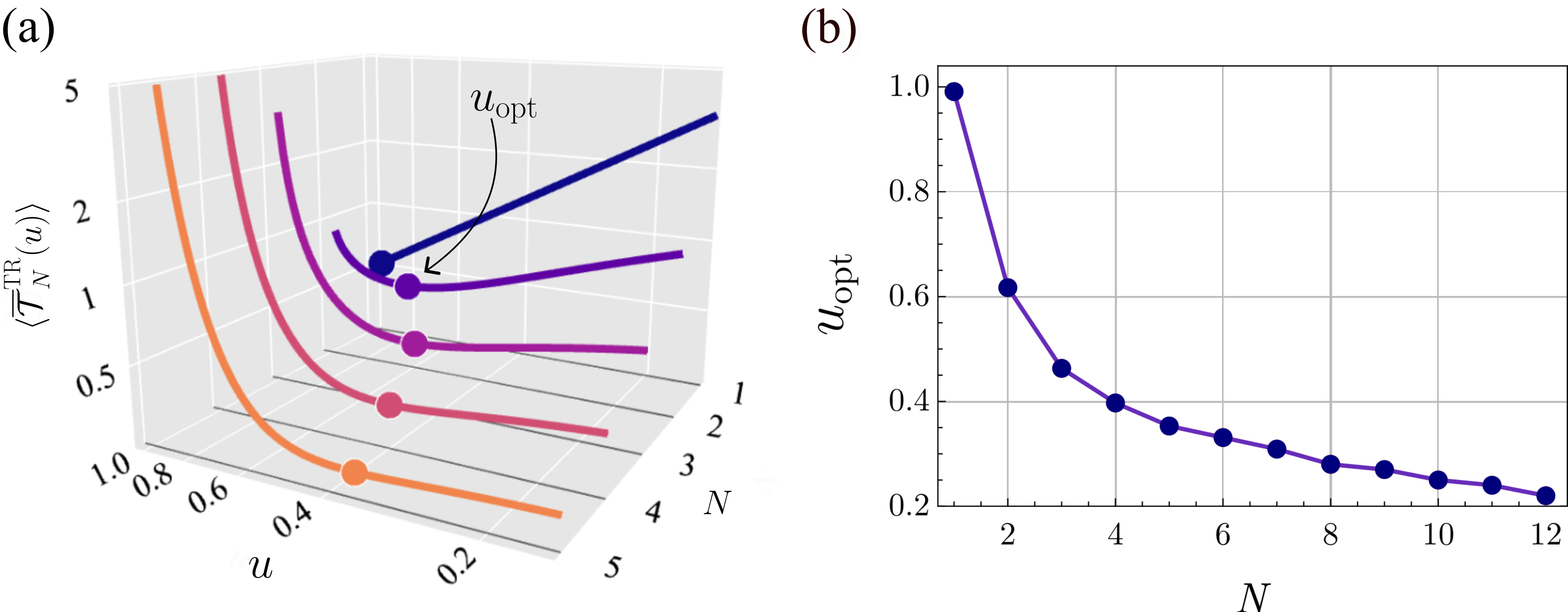}
     \caption{Panel (a) shows the variation of the non-dimensionalized MFPT $\langle   \overline{\mathcal{T}}_N^{\text{TR}}(u) \rangle$ as a function of $u=x_0/L$ for various values of $N$ as in \eref{mfpt-diff}.  For any values of $N>1$, the curves show non-monotonic behaviour with respect to $u$ with the optimal MFPT being at $u=u_{\text{opt}}$.
     Panel (b) shows the variation of the point $u_{\text{opt}}$  with respect to $N$. 
     }
     \label{fig2}
 \end{figure*}
Note that even before doing any calculation, one can predict that the dimensionless MFPT under TR should be a function of the parameters $u$ and $N$ only, so that 
\begin{align}
 \langle   \overline{\mathcal{T}}_N^{\text{TR}}(u) \rangle= \frac{D}{x_0^2}\langle \mathcal{T}_N^{\text{TR}}(L,x_0) \rangle=\mathcal{F}(u,N),   
\end{align}
with $\mathcal{F}(u,N)$ being the scaling function. Substituting the relevant observables for the single searcher as obtained in Eq. (\ref{surv-nd})-(\ref{jlnd}) into \eref{mfpt}, we obtain the exact expression of the scaling function $\mathcal{F}(u,N)$ given by
\begin{widetext}
\begin{align}
\mathcal{F}(u,N)
&=\frac{1}{u^2}\left(\frac{ \int_0^\infty dt~ \left[\frac{2}{\pi}\sum _{n=1}^{\infty} \left(\frac{1-(-1)^n }{  n} \right)\sin \left(n\pi u\right) e^{-n^2 \pi ^2 t}\right]^N}{N\int_0^\infty dt~ \left[2 \pi  \sum _{n=1}^{\infty} n \sin  \left(n\pi u\right) e^{-n^2 \pi ^2  t}\right] \left[\frac{2}{\pi}\sum _{n=1}^{\infty} \left(\frac{1-(-1)^n }{  n} \right)\sin \left(n\pi u\right) e^{-n^2 \pi ^2 t}\right]^{N-1}}\right). \label{mfpt-diff}
\end{align}
\end{widetext}
Through the above change of variables, we have recast the dimensionless mean first-passage time (MFPT), $\langle \overline{\mathcal{T}}_N^{\text{TR}}(u) \rangle$, as a function of a single dimensionless parameter $u \in (0,1)$, effectively eliminating dependence on the original parameters $L$, $x_0$, and $D$. While the expression in Eq. \eqref{mfpt-diff} does not admit a closed-form solution for arbitrary $N$, it can be evaluated numerically with high precision using Mathematica. In what follows, we examine in detail the behavior of the scaled MFPT—i.e., the scaling function $\mathcal{F}(u, N)$—as a function of both $u$ and $N$, and highlight the physical insights that emerge from this analysis.


\subsection{Optimization with respect to threshold} \label{var-threshold}
In canonical resetting frameworks, it is well established that the MFPT can be optimized by tuning the resetting frequency \cite{evans_diffusion_2011,pal_first_2017}. Unlike these externally timed protocols, the TR mechanism induces \textit{event-driven} resets—triggered when a searcher reaches a prescribed spatial threshold. In this setup, with fixed $x_0$, varying the threshold position $L$ allows control over the effective resetting rate. For instance, in the limit $L \to \infty$, resetting becomes negligible, as the threshold is rarely reached. Conversely, when $L \to x_0$, the threshold is close to the starting point, resulting in frequent resets. In terms of the dimensionless parameter $u = x_0/L$, the limit $u \to 1$ corresponds to \textit{high resetting frequency}, while $u \to 0$ captures the \textit{low frequency} regime. This naturally raises the question: can varying $u$ in the TR framework lead to an optimal MFPT, akin to externally driven resetting schemes? To explore this,  we plot the analytical result for MFPT in \eref{mfpt-diff} as a function of $u$ in Fig. \ref{fig2}(a) for various values of $N$. Remarkably, for all $N \geq 2$, the MFPT exhibits a clear minimum at an optimal value of $u$. However, this optimization is absent in the single-searcher case ($N = 1$). In what follows, we analyze these two cases in detail.\\

\noindent
\textbf{\textit{Single diffusive searcher $(N=1)$}:} For a single diffusive searcher, the MFPT can be computed by setting $N=1$ in \eref{mfpt-diff} to obtain
\begin{align}
    \mathcal{F}(u,N=1) &=\frac{1}{u^2}\left(\frac{  \frac{2}{\pi}\sum _{n=1}^{\infty} \left(\frac{1-(-1)^n }{  n} \right)\sin \left(n\pi u\right) \int_0^\infty dt~e^{-n^2 \pi ^2 t}}{ 2 \pi  \sum _{n=1}^{\infty} n \sin  \left(n\pi u\right) \int_0^\infty dt~e^{-n^2 \pi ^2  t} }\right), \nonumber\\
     &=\frac{\pi}{2u^2}\left(\frac{2\sum _{n=1}^{\infty} \left(\frac{1-(-1)^n }{  n^3\pi^3} \right)\sin \left(n\pi u\right)}{  \sum _{n=1}^{\infty} \frac{\sin(n\pi u)}{n}}\right),
\end{align}
which can be simplified by noting that in the numerator only odd $n$ will contribute and by noting the following identities \cite{gradshteyn2014table}
\begin{align}
    \sum_{n=0}^{\infty} \frac{\sin\big((2n+1)x\big)}{(2n+1)^3}
&= \frac{\pi^2 x}{8} - \frac{\pi x^2}{8}, \quad 0 < x < \pi \\
\sum_{n=1}^{\infty} \frac{\sin(n x)}{n}
&= \frac{\pi - x}{2}, \quad 0 < x < 2\pi.
\end{align}
Further simplification leads to
\begin{align}
    \langle\overline{\mathcal{T}}_1^{\text{TR}}(u) \rangle = \mathcal{F}(u,N=1)=\frac{1}{2u}. \label{mfpt-n1}
\end{align}
\noindent
From \eref{mfpt-n1} it is evident that the MFPT for a single diffusive searcher decreases as $u^{-1}$ as also shown in \fref{fig2}(a). It is the lowest when $L\to x_0$ ($u\to 1$) where $ \langle \overline{\mathcal{T}}_1^{\text{TR}}(u=1) \rangle \rangle =1/2$. Clearly here $u \to 1$ is the optimal point ($u_{\text{opt}}$) as also observed in \fref{fig2}(b). Physically, such behavior can be explained in the following way:  The threshold effectively biases the searcher's motion towards the target by resetting it to $x_0$ whenever it goes away from the target. Keeping $x_0$ fixed, as one decreases $L$, therefore increasing $u$, the chances of the searchers wandering away from the target also diminish, resulting in a lower MFPT.\\\\
\textbf{\textit{Multiple diffusive searchers $(N\ge 2)$}:}
The MFPT exhibits quite distinct features for $N\ge 2$ than that for $N=1$. In here, the MFPT curves show a non-monotonic behaviour with respect to $u$ as seen in \fref{fig2}(a). In the following, we delve further to discuss various limits of the curves and quantify the optimization features. The limit $u\to 0$ is just the reset-free case where the threshold is kept at infinity (assuming $x_0$ to be fixed). In this case, the MFPT $\langle \overline{\mathcal{T}}_N^{\text{TR}}(u\to 0) \rangle$ is just the mean fastest first passage time out of $N$ searchers to reach the target in the absence of the threshold. The MFPT in this limit (\aref{up-diff}) is found to be
\begin{align}
    \langle \overline{\mathcal{T}}_N^{\text{TR}}(u\to 0) \rangle=\frac{1}{2} \int_0^\infty dy ~y~\left[\text{erf}\left(\frac{1}{y}\right)\right]^N, \label{lim-u-0}
\end{align}
which is infinite for $N=1,2$ but becomes finite only for $N\ge 3$. However, as one increases $u$ the MFPT starts to decrease, showing an optimum at some intermediate value $u=u_{\text{opt}}$ (as shown in \fref{fig2}(a)) before increasing again as $u$ approaches unity.  Here,  $u_{\text{opt}}$  can be thought of as the analogue of the optimal resetting rate in the traditional resetting set-up \cite{evans_diffusion_2011}. In \fref{fig2}(b) we show the variation of $u_{\text{opt}}$ with respect to $N$. For $N=1$ we find $u_{\text{opt}}=1$ since the lowest MFPT occurs only at $u=1$, as evident from \eref{mfpt-n1}. However, for any other values of $N\ge 2$ we find $0<u_{\text{opt}}<1$. While the MFPT without TR ($u \to 0$) for diffusive searchers remains finite for $N\ge 3$ (\aref{up-diff}), further reduction can be observed by tuning $u$. This is clearly due to a finite threshold that prevents the searchers from wandering off the target. In turn, when $u\to 1$, the MFPT increases again for $ N\ge 2$ (in contrast to $N=1$). Since the searchers start very close to the threshold in this limit, the splitting probability $\epsilon_L$ to hit the threshold is significantly higher and thus, the TR mechanism will impede the overall search process. With increasing $N$, the probability of hitting the threshold increases even more and eventually, the probability of finding the target diminishes, leading to a steady increase in the MFPT. We refer to section \ref{speed-up} for a quantitative analysis of the efficiency gained through optimal threshold resetting with respect to the reset-free process ($u\to 0$).

\begin{figure*}
     \centering
     \includegraphics[width=17cm]{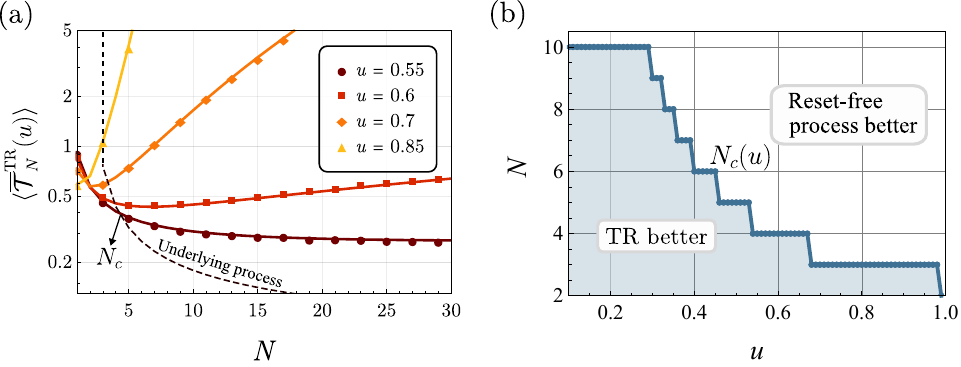}
     \caption{ Panel (a) shows the variation of the non-dimensionalized MFPT  with the number of searchers $N$ for various values of $u$. The solid line represents the analytical results (\eref{mfpt-diff}) and the markers represent results from simulation. The dashed line corresponds to the underlying process (\textit{i.e.} $u\to 0$). Note that, for a fixed $u$, when the solid curves with $u \ne 0$ lie below the dashed curve, MFPT with TR turns out to be a favourable strategy than the underlying process. The intersection point where these two curves cross each other is denoted by $N_c(u)$. This is the critical number of searchers, below which TR serves as a better strategy than the underlying process. In panel (b) we show the variation of $N_c(u)$ with $u$ (the solid circles).  This can be interpreted in the following way: for a fixed range of $u$ when $N \le N_c(u)$, TR helps to expedite the collective search process (shown by the shaded region). For instance, for $0.635 \leq u \leq 0.97$, one can always find that $N<N_c=3$ is TR beneficial. }
     
     \label{fig3}
 \end{figure*}

\subsection{Optimization with respect to number of searchers} \label{var-searchers}
Besides the optimization with respect to the threshold, the MFPT also shows rich optimization features with respect to the number of searchers $N$. In this section, we discuss the variation of the scaled MFPT (as in \eref{mfpt-diff}) with respect to $N$. It turns out that, for a fixed $u$, there exist two distinct kinds of optimisation. In the first case, we note an optimization rendered by collective search in comparison to the underlying reset-free process -- we find a critical number of searchers $N_c(u)$ such that for any $N\le N_c(u)$, the MFPT with TR is lower than that of the underlying reset-free process. In the second case, we find that there exists an optimal number of searchers $N_{\text{opt}}(u)$ for which the MFPT under TR can be made the lowest. In the following, we  elaborate on both of these cases separately.

\subsubsection{\textbf{Critical number of searchers}}
To demonstrate the efficiency gained by the collective search, we turn our attention to \fref{fig3}(a), which shows the variation of MFPT $\langle   \overline{\mathcal{T}}_N^{\text{TR}}(u) \rangle$ with respect to the number of searchers for different $u$.  The dashed line represents the MFPT as a function of $N$ for the underlying reset-free process ($u\to 0$) as in \eref{lim-u-0}. The dashed line intersects the solid lines (MFPT with TR) at a critical number $N_c(u)$ for each $u$ (for instance, $N_c(u=0.55)=4$ as shown in \fref{fig3}(a)). For $N<N_c(u)$, we always find that the MFPT under TR is lower than the underlying ($u \to 0$) process. 

The critical number of searchers, $N_c(u)$, varies with the parameter $u$, and this dependency is captured by the solid circles in Fig. \ref{fig3}(b), which delineates the phase boundary.  It should be noted that $N_c(u)$ is same for a range of $u$ as shown by the plateaus in \fref{fig3}(b). The shaded region in the diagram identifies the regime where threshold resetting (TR) leads to a more efficient search compared to the reset-free case. This phase diagram provides practical insight: for a fixed $u$, one can determine the minimum number of searchers $N < N_c(u)$ required to benefit from TR. Conversely, for a fixed number of searchers, the diagram helps identify critical threshold positions—encoded in $u$—that yield a speed-up in the search process.


\subsubsection{\textbf{Optimal number of searchers}}
\begin{figure}[h!]
    \centering
    \includegraphics[width=8cm]{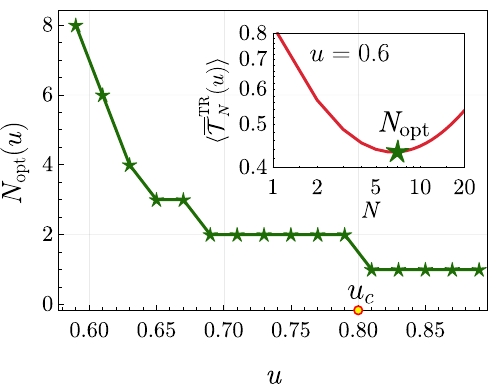}
    \caption{ Variation of the optimal number of searchers $N_{\text{opt}}(u)$ where the MFPT with TR is the lowest with respect to $u$. As a representative case, in the inset, we show $N_\text{opt}(u=0.6)$ marked by the star. The critical $u_c \approx 0.8$, beyond which the collective search becomes detrimental, is marked by an open circle.}
    \label{fig33}
\end{figure}

We also observe an optimization under TR with respect to the number of searchers. From \fref{fig3}(a), it is observed that the MFPT shows non-monotonic dependence on $N$ for certain choices of $u$.
For instance, consider the MFPT plot with $u=0.6$ as also shown in the inset of \fref{fig33}. In this case, there exists an optimal number of searchers $N_{\text{opt}}(u=0.6)=7$ for which the MFPT is the lowest. Mathematically, for a fixed value of $u$ at $N=N_{\text{opt}}(u)$ the MFPT is lowest so that one has
\begin{align}
   \left. \frac{d \langle   \overline{\mathcal{T}}_N^{\text{TR}}(u) \rangle}{d N} \right|_{N=N_\text{opt}}=0.
\end{align}
 \fref{fig33} shows the variation of $N_{\text{opt}}(u)$ with respect to $u$, where we note that beyond a critical value of $u=u_c \approx 0.8$,  the optimal number of searchers is fixed to one. This in turn implies that, beyond $u>u_c$, the MFPT will grow monotonically with $N$, making the collective search detrimental. However, for $u<u_c$, the MFPT can be minimized for a suitable choice of $N_{\text{opt}}(u)\ge 2$.\\
 
\textbf{\textit{Asymptotic behaviour}:} As the number of searchers grows very large such that $N \to \infty$, the MFPT is not guaranteed to be finite for any arbitrary choice of $u$. For instance, consider the case when the searchers start very close to the threshold so that $x_0 \to L$ or equivalently, $u\to 1$. In that case, for $N\gg 1$, at least one of the $N$ searchers will hit the threshold in a much shorter time span than hitting the target. Naturally, the frequency of resetting will be much higher. Moreover, increasing $N$ will further enhance the probability of hitting the threshold, causing further delay in the target search. On the other hand, when the searchers start very close to the target such that $u \to 0$, increasing the number of searchers should further diminish the MFPT. To find the exact asymptotic variation of the MFPT for large $N$, we express the MFPT (\eref{mfpt-diff}) in  the following form,
\begin{align}
    \mathcal{F}(u,N)=\frac{1}{u^2}\frac{I_N(u)}{J_N(u)},
\end{align}
where,
\begin{align}
    I_N(u)&= \int_0^\infty dt~ \left[\frac{2}{\pi}\sum _{n=1}^{\infty} \left(\frac{1-(-1)^n }{  n} \right)\sin \left(n\pi u\right) e^{-n^2 \pi ^2 t}\right]^N,\\
    J_N(u)&=N\int_0^\infty dt~ \left[2 \pi  \sum _{n=1}^{\infty} n \sin  \left(n\pi u\right) e^{-n^2 \pi ^2  t}\right] \left[\frac{2}{\pi}\sum _{n=1}^{\infty} \left(\frac{1-(-1)^n }{  n} \right)\sin \left(n\pi u\right) e^{-n^2 \pi ^2 t}\right]^{N-1}.
\end{align}
In the limit of $N \to \infty$ the functions $I_N(u)$ and $J_N(u)$ have the following asymptotic form (i.e., for fixed $u$ and $N\to \infty$) (see \aref{appd} for derivation)
\begin{align}
     I_N(u) \sim \frac{\min(u, 1-u)^2}{4 \ln N},~~~\text{and}~~ \begin{array}{l}
         J_N(u) \approx \left\{\begin{array}{lll}
 1,  ~~  \text{ when} ~ u \to 0,  \vspace{0.3cm} \\
     2\sqrt{\pi}N^{3/2} \left[\frac{2(1-u)^2N}{\pi e}\right]^{N/2}, ~~  \text{ when} ~ u \to 1.
          \end{array}\right.\end{array}
\end{align}
Similar results have already been studied earlier in the context of extreme value statistics \cite{weiss1983order,krapivsky2010maximum,linn2022extreme}. However, for the sake of completeness, we re-derive them starting from our general result \eref{mfpt-diff} in the \aref{appd}. Eventually, the asymptotic behaviour of $\mathcal{F}(u,N)$ up to leading order in $N$, is governed by
\begin{align} 
     \begin{array}{l}
         \mathcal{F} (u,N) \approx \left\{\begin{array}{lll}
 \frac{1}{ 4\ln N},  ~~  \text{ when} ~ u \to 0,  \vspace{0.3cm} \\
     \exp[- N \ln C(u,N)] , ~~  \text{ when} ~ u \to 1.
          \end{array}\right. \end{array},
\end{align}
where $C(u,N)=(1-u)\sqrt{\frac{2N}{\pi e}}$. Quite interestingly, 
the dependence of the MFPT on $N$ exhibits a drastic transition across a critical threshold $u^{BL} = 1 - \sqrt{\frac{\pi e}{2N}}$, which defines a boundary layer of thickness $\sim 1/\sqrt{N}$ near the starting point. If the threshold lies within this boundary layer ($u^{BL} < u < 1$, yielding $C < 1$), the MFPT actually increases with increasing $N$. Physically, when the starting position is in such extreme proximity to the threshold, a larger $N$ drastically amplifies the rate of threshold collisions. This frequent hitting effectively traps and localizes the particles near their initial position, preventing them from exploring the wider domain and severely delaying their arrival at the target. Conversely, if the threshold sits just outside this boundary layer ($u < u^{BL}$, yielding $C > 1$), the MFPT decreases as $N$ grows. In this regime, the starting point is sufficiently distant to avoid this intense localization effect; instead, increasing $N$ broadens the dynamics, allowing a greater fraction of particles to bypass the threshold entirely and proceed directly to the target, thereby accelerating the overall process. The existence of such a boundary layer in this problem is indeed intriguing and needs further exploration in future.

\begin{figure}
    \centering
    \includegraphics[width=16cm]{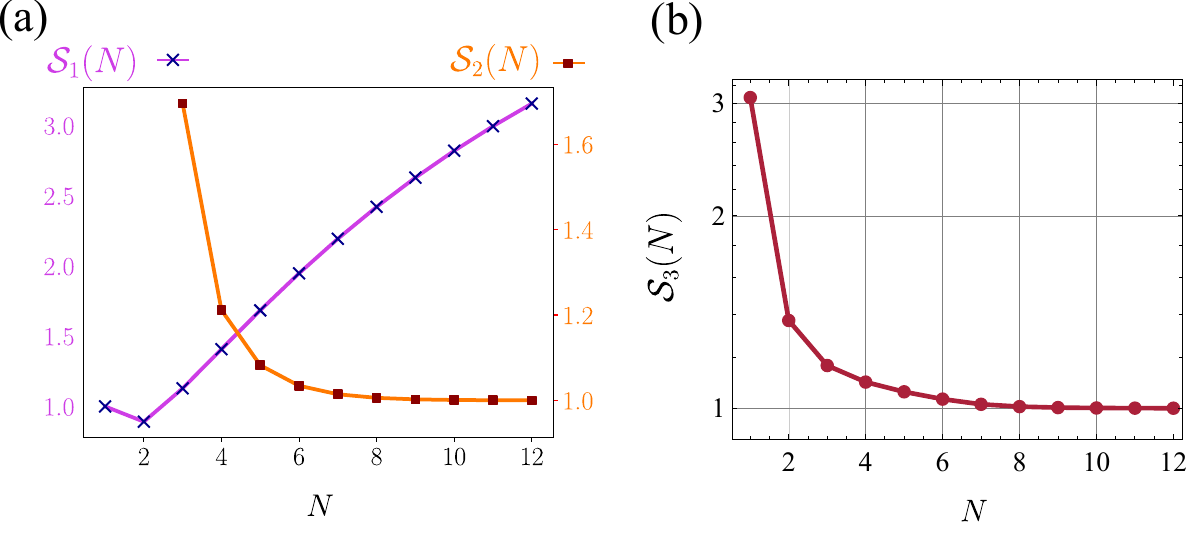}
    \caption{ Variation of the optimal speed-up gained under threshold resetting mechanism with $N$. (a) For the function values $>1$ (marked on the $y$-axis), TR optimizes the search process. The blue cross shows the variation of $\mathcal{S}_1(N)$, which quantifies the speed-up with gained with multiple searchers compared to a single searcher. The quantity $\mathcal{S}_2(N)$ shown by the red squares amounts to the speed-up gained with TR in comparison to the underlying reset-free search process.  (b) The parameter  $\mathcal{S}_3(N)$ quantifies the relative speed-up 
    between
    a TR process and the standard PR process, both conducted at their respective optimality. It is noted from $\mathcal{S}_3(N)>1$ that TR always outperforms PR in a collective diffusive search. }
    \label{fig6}
\end{figure}

\subsection{Optimal speed-up due to collective TR}
\label{speed-up}
In this section, we aim to quantify the efficiency gain achieved by employing threshold resetting (TR) in a system of $N$ non-interacting diffusive searchers. Specifically, we assess the speed-up provided by the collective TR mechanism by evaluating how much it reduces the mean first-passage time (MFPT) compared to alternative search strategies. To do so, we focus on the \textit{minimum MFPT} obtained under TR, which occurs at the optimal threshold parameter $u = u_{\text{opt}}$, and denote it as $\langle \overline{\mathcal{T}}_N^{\text{TR}}(u_{\text{opt}}) \rangle$. This represents the best-case performance of the TR protocol for a given number of searchers $N$. We then compare this optimized TR-MFPT with two relevant baseline MFPTs derived from conventional search setups, each optimized in its own right. These comparisons define two complementary notions of \textit{optimal speed-up}, as detailed below. These measures allow us to systematically characterize the advantage offered by threshold resetting.

\subsubsection{Speed-up relative to the collective TR strategy}
The speed-up function $\mathcal{S}_1(N)=\frac{ \langle   \overline{\mathcal{T}}_{N=1}^{\text{TR}}(u_\text{opt}) \rangle}{\langle   \overline{\mathcal{T}}_N^{\text{TR}}(u_\text{opt}) \rangle}$ quantifies the efficiency gained with optimal TR by employing multiple searchers ($N\ge 2$) compared to a single searcher ($N=1$). The variation of this quantity with respect to $N$ is shown by the blue crosses in \fref{fig6} with its values marked in the left $y$-axis of the plot. In general, we observe a speed-up for any $N\ge 3$. With increasing $N$ further, the speed-up also enhances significantly. 

 \subsubsection{Speed-up relative to the reset-free process} The function $\mathcal{S}_2(N)=\frac{ \langle   \overline{\mathcal{T}}_{N}^{\text{TR}}(u=0) \rangle}{\langle   \overline{\mathcal{T}}_N^{\text{TR}}(u_\text{opt}) \rangle}$ quantifies 
the speed-up achieved by the TR mechanism relative to the baseline reset-free process which can be obtained by setting $L \to \infty$, or equivalently, $u \to 0$ (depicted by the red squares in \fref{fig6}, with corresponding values indicated on the right $y$-axis). As previously discussed, for $N \le 2$, the MFPT in the absence of resetting is divergent, whereas TR renders it finite—highlighting an immediate speed-up. For $N \ge 3$, although the reset-free MFPT is already finite, introducing optimal TR still results in a further reduction. This improvement is evident in \fref{fig6}, where the speed-up exceeds unity for $N \ge 3$. However, as $N$ continues to increase, the benefit from TR gradually diminishes, and the speed-up approaches unity—indicating that TR offers no additional advantage in the large-$N$ limit.

 \subsubsection{Speed-up relative to Poissonian resetting}
 Let us now ask whether TR  is a better strategy than the standard Poissonian resetting (PR) \cite{biroli2023critical} in reducing the MFPT of a collective diffusive search. Particularly, we compare the MFPT of a collective diffusive search at the optimal resetting rate $r_\text{opt}$ against the MFPT by TR at the optimal value of $u=u_{opt}$ as shown in \fref{fig2}(b). This comparison can be captured by the speed-up parameter $\mathcal{S}_3(N)=\frac{ \langle   \overline{\mathcal{T}}_{N}^{\text{PR}}(r_\text{opt}) \rangle}{\langle   \overline{\mathcal{T}}_N^{\text{TR}}(u_\text{opt}) \rangle}$, where the MFPT of a collective diffusive search by $N$ particles with resetting at a rate $r$ is denoted by $\langle   \overline{\mathcal{T}}_{N}^{\text{PR}}(r) \rangle$. The quantity $\langle   \overline{\mathcal{T}}_{N}^{\text{PR}}(r) \rangle$ can be found following Biroli \textit{et al} \cite{biroli2023critical} and is given by
 \begin{align}
     \langle   \overline{\mathcal{T}}_{N}^{\text{PR}}(r) \rangle=\frac{\int_0^\infty dt e^{-rt} \left[\text{erf}\left(\frac{1}{\sqrt{4 t}}\right)\right]^N}{1-r\int_0^\infty dt e^{-rt} \left[\text{erf}\left(\frac{1}{\sqrt{4 t}}\right)\right]^N},
 \end{align}
where both the MFPT and the resetting rate are non-dimensionalised by the scaling factor $x_0^2/D$. It should be noted that the study \cite{biroli2023critical} shows that for $N>6$, PR cannot accelerate the process. However, for $N\le 6$ there exists an $r_\text{opt}$ at which the MFPT with PR is the lowest. In \fref{fig6}(b) we show the variation of the parameter $\mathcal{S}_3(N)$ for several values of $N$. Interestingly, we find that for any number of searchers, $\mathcal{S}_3(N)$ remains greater than unity. This result in turn implies that the optimal TR always beats optimal PR in a collective search process. The amount of speedup is highest for $N=1$ and starts to reduce as $N$ increases further and saturates to unity. Such behaviour could also be understood from the fact that, for $N>6$, PR does not accelerate the search \cite{biroli2023critical} such that $r_\text{opt}=0$. On the other hand, from \fref{fig2}(b) it is observed that with increasing $N$, the value of $u_\text{opt}$, while approaching zero, stays finite, yielding a minimum MFPT, thus effectively rendering the speed-up. For large values of $N$, the underlying reset-free search process becomes the best possible strategy for both the TR and PR, leading to $\mathcal{S}_3(N)=1$.

 \subsubsection{Speed-up relative to reflective boundary without resetting}
 \begin{figure}[t]
     \centering
     \includegraphics[width=16cm]{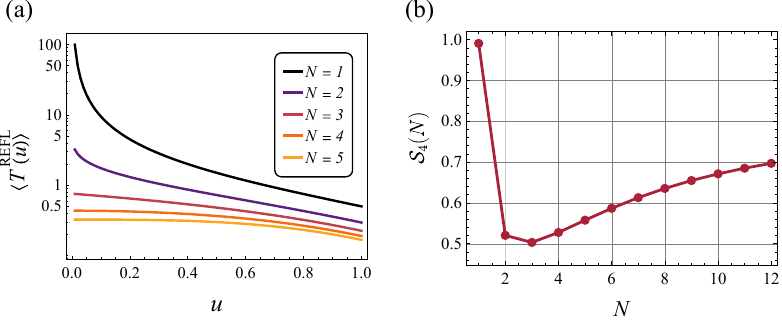}
     \caption{(a) Variation of the MFPT with a reflective boundary replacing the threshold (\eref{ref-mfpt}). Note that $u=1$ is the optimal value for any value of $N$. (b) Variation of the optimal speed-up $S_4(N)$ with $N$. The value of $S_4(N) \le 1$ implies that the reflective search is always better than the optimal TR.} 
     \label{s4n}
 \end{figure}

 Let us now consider the same setup as discussed in this paper, except that the threshold boundary is replaced by a reflecting one. In here, we have $N$ Brownian particles which start from $x_0$ and there is a reflecting boundary at $x=L$. This is a completely reset-free search process and the search is completed whenever any one of the particles hits the target at $x=0$.
 Since the system is again bounded, the MFPT for the collective diffusive search is always finite in this case, even in the absence of resetting. We ask whether the optimal TR can perform better than the optimal non-reset reflective search. In here, the particles do not generate correlations unlike TR where the searchers become dynamically coupled through simultaneous resetting events, giving rise to a genuinely collective search process \cite{majumdar2026dynamically,biswas2025target}.

 Using standard techniques of first-passage processes, we compute the MFPT $\langle   \overline{\mathcal{T}}_{N}^{\text{REFL}}(u) \rangle$ for $N$ diffusing particles in the presence of a reflecting boundary \cite{redner2001}
\begin{align}
    \langle   \overline{\mathcal{T}}_{N}^{\text{REFL}}(u) \rangle=\int_0^\infty dt \frac{1}{u^2} \left( \frac{2}{\pi} \sum_{n=0}^{\infty} \frac{1}{n + \frac{1}{2}} \sin\left[ \left(n + \frac{1}{2}\right) \pi u \right] e^{-\left(n + \frac{1}{2}\right)^2 \pi^2 t} \right)^N, \label{ref-mfpt}
\end{align}
where the MFPT is scaled by a factor of $x_0^2/D$. In \fref{s4n}(a) we show the variation of $ \langle   \overline{\mathcal{T}}_{N}^{\text{REFL}}(u) \rangle$ with respect to $u$ for several values of $N$. It is found that for any values of $N$, the optimal search is performed when $u=1$. This is physically evident since for $u=1$ or $L=x_0$, the particles have zero chances of spending time away from the target. In \fref{s4n}(b) we show the variation of the optimal speed-up parameter $\mathcal{S}_4(N)=\frac{ \langle   \overline{\mathcal{T}}_{N}^{\text{REFL}}(u=1) \rangle}{\langle   \overline{\mathcal{T}}_N^{\text{TR}}(u_\text{opt}) \rangle}$ as a function of $N$. It turns out that for $N=1$ both the optimal TR and optimal reflective search have the same MFPT such that $\mathcal{S}_4(N)=1$. However, for any values of $N>1$ we see that $\mathcal{S}_4(N)<1$ such that reflective search performs better than the optimal TR. Physically, this can be understood as follows: For $N=1$ we have $u_{\text{opt}}=1$ so the particle is reset directly to the threshold, effectively mimicking a reflective boundary. Thus MFPT for both the cases are equal, and we have $S_4(N=1)=1$. On the other hand, consider the limit of very large $N$ where $u_\text{opt} \to 0$, as evident from \fref{fig2}(b). In that case, $L \to \infty$ or absence of TR is the best possible scenario; in other words, the particles have a finite chance of spending time away from the target as well. In contrast, keeping a reflective boundary at $L=x_0$ makes the particles reach the target faster, thereby improving performance.

Finally, it is important to emphasize that the threshold-resetting dynamics and the corresponding reset-free process with a reflecting boundary represent fundamentally different search models. In the latter, the walkers evolve independently throughout the search, whereas under TR the searchers become dynamically coupled through simultaneous resetting events, giving rise to a genuinely collective search process \cite{majumdar2026dynamically}. Henceforth, the key objective of the paper is not to compare competing search strategies, but to understand and optimize the performance of the TR protocol itself with respect to the threshold position and the number of searchers.

\section{Cost function} \label{costs}
A key challenge in any optimization problem is managing the cost involved in achieving a globally optimal outcome for a given observable. In the context of traditional stochastic resetting, where the dynamics are reset after random waiting times, the notion of cost (thermodynamic and dynamic) has been previously studied in various works \cite{de2020optimization,sunil2023cost,sunil2024minimizing,tal2020experimental,tal2025smart,olsen2024thermodynamic,gupta2020work,pal2023thermodynamic}. 
 In TR optimization problems, a cost can be incorporated into an objective function to balance between faster completion times and the penalty associated with each resetting event. Thus, the cost function $C_N(u)$ for our threshold resetting (TR) setup, following the formulation originally proposed in \cite{biswas2025target}, can be written as
\begin{align}
    C_N(u)=\langle   \overline{\mathcal{T}}_N^{\text{TR}}(u) \rangle + \beta N \langle{\mathcal{N}}_{\text{TR}} (N,u) \rangle, 
\end{align}
where $\langle{\mathcal{N}}_{\text{TR}} (N,u) \rangle$ is the mean number of resetting events before the first passage with TR and $\beta$ is some arbitrary constant that has the dimension of the inverse of time. Physically $\beta$ can be thought of as some constant cost per resetting event. The existence of optimality of $C_N$ for certain choice of system parameters would in principle imply that the MFPT can sufficiently be optimized at those parameter values without paying too much cost for resetting events.

\begin{figure}
    \centering
    \includegraphics[width=8cm]{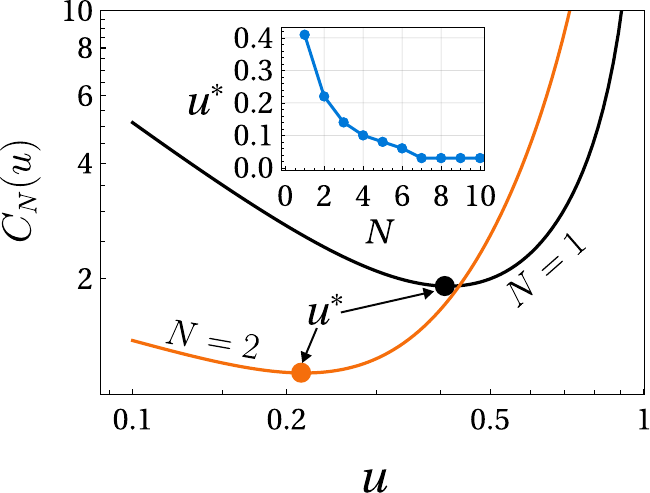}
    \caption{Variation of cost function as defined in \eref{cost} for diffusive search with TR with respect to $u=x_0/L$. Note that, unlike the MFPT, the cost function shows non-monotonic behaviour even for $N=1$. The variation of optimal $u^*$ where cost is minimized is shown for various values of $N$ in the inset. }
    \label{diff-cost}
\end{figure}

The quantity $\langle{\mathcal{N}}_{\text{TR}} (N,u) \rangle$ can be easily found from a trajectory-based analysis. Suppose there are $k$ number of resetting events in a single chosen trajectory until the first passage. Notice that the probability that a resetting event will occur is nothing but the splitting probability to the threshold i.e., $\epsilon_L (N,u)$. As there are $k$ resetting events thus the particular trajectory occurs with probability $[\epsilon_L (N,u)]^k \times \epsilon_0 (N,u)$. The quantity $\epsilon_0 (N,u)=1-\epsilon_L (N,u)$ is multiplied to take care of the probability of finding the target after the $k^{th}$ resetting event. The mean number of resetting events till the first passage then can be found as
\begin{align}
    \langle{\mathcal{N}}_{\text{TR}} (N,u) \rangle &= \sum_{k=0}^\infty k [\epsilon_L (N,u)]^k \times \epsilon_0 (N,u) =\frac{ \epsilon_L (N,u)}{ \epsilon_0 (N,u)}.
\end{align}
Finally, we can find the cost function as
\begin{align}
     C_N(u)=\langle   \overline{\mathcal{T}}_N^{\text{TR}}(u) \rangle + \beta N  \left[\frac{ \epsilon_L (N,u)}{ \epsilon_0 (N,u)}\right].\label{cost}
\end{align}
In \fref{diff-cost}(a) we show the variation of the cost function for diffusive searchers both with respect to $u$. Note that the cost function shows optimal behaviour with respect to $u$ even for a single diffusive searcher. This is because, although the MFPT is optimal while $u\to 1$, the mean number of resetting diverges in this limit. Variation of the optimal value of $u^*$ where the cost becomes the lowest is shown in the inset of \fref{diff-cost}(a) for different choices of $N$. It is observed that $u^* \to 0$ as number of searchers $N$ increases. This is because the MFPT at larger values of $N$ is already optimal for the reset-free process (recall that $u_{\text{opt}}\to 0$ as $N \gg 1$). Thus, the introduction of resetting (or a finite value of $u$) only increases the contribution from the second term in \eref{cost} without any substantial improvement in the MFPT. Concluding, this analysis highlights how an optimal resetting strategy can effectively mitigate larger, more detrimental outcomes. This approach can be seen as an optimization technique, balancing the cost of resetting against the potential savings in long-term performance, ensuring that the system operates in the most efficient manner possible in threshold-driven events.

 The cost function used in the paper is one simple possible choice to study how to optimize the MFPT while simultaneously balancing
the resetting cost. Another realistic and
experimentally relevant choice for the cost function would be to consider the amount of distance a particle had to travel before being reset. This idea has been explored in several earlier works in the context of standard resetting \cite{tal2020experimental,besga2020optimal,pal_search_2020,pal2019invariants,biroli2026experimental}. Incorporating such aspects to the TR is left for future studies.

\section{Discussion and outlook} \label{disc}
To summarize, in this work, we have employed the strategy of threshold resetting originally developed in \cite{biswas2025target} to investigate the target search properties by $N$ non-interacting diffusive searchers. Besides revisiting the formalism presented in \cite{biswas2025target}, we also provide an alternate method based on the renewal theory of random variables to extract the first passage statistics of the process. Then we elaborate the formalism with the paradigmatic example of diffusive searchers on one dimension. In our set-up, the origin $x=0$ is the target, and the boundary at $x=L$ serves as the resetting threshold.  We consider a resetting protocol in which all searchers are reset simultaneously to the initial position whenever any one of them reaches the threshold. We found that akin to resetting rate in the externally driven resetting protocol, under TR one can tune the dimensionless parameter $u=x_0/L$ to modulate the resetting frequency. For a single diffusive searcher, we find the MFPT monotonically decreases with $u$ with $u=1$ being the optimal point. Quite interestingly, when the search is conducted by more than one diffusive searcher, we find a rich optimization behavior with respect to $u$. A preliminary version of this result was reported in \cite{biswas2025target}. Here, we present a more detailed and comprehensive analysis of collective diffusive search. In particular, we demonstrate that the system exhibits rich optimization behavior with respect to the threshold and the number of searchers. Combining both the results, we found a parameter space in $u-N$ plane that separates the regimes where TR is helpful from those where it is not. We see for a fixed $u$ there exists a critical number of searchers $N_c(u)$ below which the search with TR will expedite the reset-free search process. 
Next, we optimize the MFPT with respect to the number of searchers ($N$), which allows us to identify an optimal population size $N_{\text{opt}}(u)$ that maximizes the efficiency of collective search. This establishes a clear bound beyond which adding more searchers no longer yields additional benefit. Additionally, we show that collective search with an optimal $u_{\text{opt}}$ perform better than the same with Poissonian resetting at an optimal rate $r_{\text{opt}}$. We further characterize the operational cost of the search, following \cite{biswas2025target}, and reveal that it exhibits a rich and nontrivial dependence on the threshold parameter. In particular, we find that the cost function admits an optimal threshold for any given number of searchers, highlighting an inherent trade-off between search efficiency and resource expenditure.


Although threshold resetting appears in a wide range of natural and engineered systems, it has received considerably less theoretical attention compared to the more conventional externally driven resetting frameworks. Further valuable extensions going beyond diffusive searchers would be the study of first-passage properties under TR for more complex dynamics and interacting agents. A natural extension would be to generalize the TR-driven search process to higher dimensions where geometric effects can play a significant role. It would also be interesting to consider scenarios with multiple targets, where competition between targets and the interplay with threshold-triggered resetting may give rise to new optimization features. Exploring these directions could provide deeper insight into the efficiency and robustness of threshold-based search strategies in more realistic settings. Beyond physics, TR has potential applications in diverse fields including economics, population dynamics, and operations research. We hope that the present work serves as a foundation for a broader exploration of threshold-driven resetting phenomena.

\section{Acknowledgement}
SNM thanks M. Biroli, S. Redner and G. Schehr for discussions on related models. AP
sincerely thanks D. Ghosh for pointing towards some relevant works. The numerical calculations reported in this
work were carried out on the Kamet cluster, which is
maintained and supported by the Institute of Mathematical Science’s High-Performance Computing Center. AB acknowledges support from USIEF, India, for the Fulbright-Nehru doctoral research fellowship.
AB and AP gratefully acknowledge research support from
the Department of Atomic Energy, Government of India. 
SNM acknowledges support from ANR Grant No. ANR23- CE30-0020-01 EDIPS. SNM and AP thank the Higgs
Center for Theoretical Physics, Edinburgh, for hospitality during the workshop “New Vistas in Stochastic Resetting” and Korea Institute for advanced Study(KIAS),
Seoul, for hospitality during the conference “Nonequilibrium Statistical Physics of Complex Systems” where several discussions related to the project took place. SNM
and AP also acknowledge the research support from the
International Research Project (IRP) titled “Classical
and quantum dynamics in out of equilibrium systems”
by CNRS, France. AP acknowledges research funding under the scheme ANRF/ARGM/2025/001623 from ANRF, India.

\section*{Appendix}
\appendix

\section{Additional steps related to Formalism II 
} \label{formalism-II-Appendix}
In this section, we present the additional steps used to derive the expressions introduced in the main text. We start by recalling Eq. (\ref{ren-2}) from the main text
\begin{align}
     \langle e^{-s \mathcal{T}_{N}^{\text{TR}}} \rangle= \frac{\langle I(\mathcal{T}_{0,N}<\mathcal{T}_{L,N}) e^{-s\mathcal{T}_{0,N}}\rangle}{1-\langle I(\mathcal{T}_{L,N}\le \mathcal{T}_{0,N}) e^{-s\mathcal{T}_{L,N}} \rangle}. \label{ren-2-Appendix}
\end{align}
The numerator can be written as
\begin{align}
    \langle I(\mathcal{T}_{0,N}<\mathcal{T}_{L,N}) e^{-s\mathcal{T}_{0,N}}\rangle
     &= \text{Pr}(\mathcal{T}_{0,N}<\mathcal{T}_{L,N}) \langle e^{-s \{\mathcal{T}_{0,N}|\mathcal{T}_{0,N}<\mathcal{T}_{L,N}\}} \rangle\nonumber\\
     &=\text{Pr}(\mathcal{T}_{0,N}<\mathcal{T}_{L,N}) \int_0^\infty dt~ f_{\mathcal{T}_{0,N}^c}(L,x_0,t) e^{-s t}
 \end{align}
where $f_{\mathcal{T}_{0,N}^c}(L,x_0,t)$ is the density associated with the conditional time $\mathcal{T}_{0,N}^c\equiv \{\mathcal{T}_{0,N}|\mathcal{T}_{0,N}<\mathcal{T}_{L,N}\}$. This density can be written as
\begin{align}
    f_{\mathcal{T}_{0,N}^c}(t)=\frac{j_{0,N}(L,x_0,t)}{\text{Pr}(\mathcal{T}_{0,N}<\mathcal{T}_{L,N})}, \label{toden-Appendix}
\end{align}
where $j_{0,N}(L,x_0,t)$ is the probability current at the target $x=0$  at time $t$ due to any one of the $N$ searchers. Using the above relation, we finally arrive at the result
\begin{align}
    \langle I(\mathcal{T}_{0,N}<\mathcal{T}_{L,N}) e^{-s\mathcal{T}_{0,N}}\rangle&=\int_0^\infty dt~ e^{-s t} j_{0,N}(L,x_0,t) \nonumber\\
    &=\widetilde{j}_{0,N}(L,x_0,s), \label{jos-Appendix}
\end{align}
with $\widetilde{j}_{0,N}(L,x_0,s)$ being the Laplace transform of $j_{0,N}(L,x_0,t)$.  Similarly, one can also find the denominator of \eref{ren-2} as
\begin{align}
    \langle I(\mathcal{T}_{L,N}\le \mathcal{T}_{0,N}) e^{-s\mathcal{T}_{L,N}} \rangle=\widetilde{j}_{L,N}(L,x_0,s). \label{jls-Appendix}
\end{align}
Thus, the generating function of the FPT can be written as
\begin{align}
       \langle e^{-s \mathcal{T}_{N}^{\text{TR}}} \rangle =\frac{\widetilde{j}_{0,N}(L,x_0,s)}{1-\widetilde{j}_{L,N}(L,x_0,s)}, \label{mgf-Appendix}
\end{align}
which is \eqref{mgf} from the main text.

\section{Equivalence between the survival probability and the currents as in \eref{surv-und}} \label{appa}
In here, we establish the relation (\ref{surv-und})
between the survival probability and the currents that was used in the main text. To this end, note that the unconditional FPT $T_N$ can be expressed in the following way (similar to \eref{renewal-1})
\begin{align}
\begin{array}{lll}
T_N=\left\{ \begin{array}{ll}
\mathcal{T}_{0,N}, ~ & ~\text{if } ~\mathcal{T}_{0,N}<\mathcal{T}_{L,N} \vspace{0.2cm}\\
\mathcal{T}_{L,N},~ &~ \text{if }~\mathcal{T}_{L,N} \leq \mathcal{T}_{0,N}  \end{array}\right.\text{ }\end{array}.
\label{renewal-2}
\end{align}
The above equation can then be written as
\begin{align}Pr
   & T_N=I(\mathcal{T}_{0,N}<\mathcal{T}_{L,N})\mathcal{T}_{0,N} + I(\mathcal{T}_{L,N} \le \mathcal{T}_{0,N})\mathcal{T}_{L,N},
\end{align}
which allows one to compute the moment-generating function in the following way
\begin{align}
    &\langle e^{-s T_N} \rangle=\langle I(\mathcal{T}_{0,N}<\mathcal{T}_{L,N}) e^{-s\mathcal{T}_{0,N}}\rangle + \langle I(\mathcal{T}_{L,N}\le \mathcal{T}_{0,N}) e^{-s\mathcal{T}_{L,N}} \rangle .
\end{align}
Now the first and second terms in the RHS of the above equation were already computed in the main text with the results given by \eref{jos} and \eref{jls}, respectively. Eventually, combining all these results the moment-generating function of $T_N$ is  given by
\begin{align}
    &\langle e^{-s T_N} \rangle=\widetilde{j}_{0,N}(L,x_0,s)+\widetilde{j}_{L,N}(L,x_0,s),
\end{align}
which same as in \eref{surv-und} in the main text. One can then employ the relation between the survival probability and the moment generating function as in \eref{surv-mgf} to obtain
\begin{align}
      \widetilde{q}_N(L,x_0,s) &=\frac{1}{s}\left[1- \langle e^{-s T_N}\rangle\right], \nonumber \\
      &=\frac{1}{s}\left[1- (\widetilde{j}_{0,N}(L,x_0,s)+\widetilde{j}_{L,N}(L,x_0,s))\right],
      \end{align}
as in \eref{surv-und} of the main text.

\section{MFPT for diffusive searchers in absence of threshold} \label{up-diff}
In this section, we recall the results for the MFPT of $N$ Brownian searchers in the presence of a single target placed at the origin. These results are known in the literature \cite{lindenberg1980lattice,mejia2011first} and here, we summarize them for completeness.
Let us recall that the single-searcher survival probability $Q(x_0,t)$ in the presence of a target at $x=0$ (in the absence of threshold) and starting from $x_0$ is a classical result from the first-passage theory and is given by \cite{redner2001,bray2013persistence}
\begin{align}
    Q(x_0,t)=\text{erf}\left(\frac{x_0}{2 \sqrt{D t}}\right).
\end{align}
For $N$ searchers, the survival probability decays faster as it gets multiplied $N$-times. The MFPT of the underlying process is therefore given by
\begin{align}
   \langle T_N(x_0) \rangle&=\int_0^\infty  dt \left[Q(x_0,t)\right]^N \nonumber\\
    &=\int_0^\infty  dt \left[\text{erf}\left(\frac{x_0}{2 \sqrt{D t}}\right)\right]^N.
\end{align}
With the substitution $y=\frac{2\sqrt{D t}}{x_0}$ we obtain the dimensionless MFPT as 
\begin{align}
     \langle \overline{\mathcal{T}}_N^{\text{TR}}(u\to 0) \rangle&=\frac{D}{x_0^2} \langle T_N(x_0) \rangle \nonumber\\
    &= \frac{1}{2} \int_0^\infty dy ~y~\left[\text{erf}\left(\frac{1}{y}\right)\right]^N. \label{diff-und}
\end{align}
The numerical value of the MFPT evaluated from the above expression is exactly equal to that obtained by setting $u\to 0$ in \eref{mfpt-diff}. However, the data obtained from \eref{diff-und} is slightly more accurate since it does not involve an infinite sum as in \eref{mfpt-diff}. From \eref{diff-und} it can be easily checked that the integral converges only when $N\ge 3$, which in turn gives a finite value of the MFPT of the underlying process, as also discussed in the main text.

\section{Asymptotic analysis of the MFPT}
\label{appd}
In this appendix, we provide the details of the asymptotic analysis of the MFPT. 
From Eq. (31) of the main text, recall that
\begin{equation}
\mathcal{F}(u,N) = \frac{I_N(u)}{u^2 J_N(u)}, \label{eq:0}
\end{equation}
where
\begin{equation}
I_N(u) = \int_0^\infty dt \left[ \frac{2}{\pi} \sum_{n=1}^\infty \frac{1-(-1)^n}{n} \sin(n\pi u) e^{-n^2\pi^2 t} \right]^N, \label{eq:1}
\end{equation}
and
\begin{equation}
J_N(u) = N \int_0^\infty dt \left[ 2\pi \sum_{n=1}^\infty n \sin(n\pi u) e^{-n^2\pi^2 t} \right] \times \left[ \frac{2}{\pi} \sum_{n=1}^\infty \frac{1-(-1)^n}{n} \sin(n\pi u) e^{-n^2\pi^2 t} \right]^{N-1}. \label{eq:2}
\end{equation}
Our goal here is to analyse the limit $u\to 1$ first for fixed $N$ and then consider the $N \to \infty$ limit. For this purpose, we set $u = 1 - \epsilon$ and take the limit $\epsilon \to 0$ for a fixed $N$.

\subsubsection{Derivation of $I_N(u)$ for arbitrary $u$}
We begin our derivation by noting that $I_N(u)$ as in Eq. \eqref{eq:1} is symmetric with respect to $u=1/2$ such that $I_N(u)=I_N(1-u)$. So we first fix the range $0\le u\le 1/2$ such that it includes the point $u=0$. Now, the term $1-(-1)^n$ evaluates to $2$ for odd $n$ and $0$ for even $n$. By setting $n = 2m + 1$, the summation simplifies to
\begin{equation}
\text{Inner sum} = \frac{4}{\pi} \sum_{m=0}^\infty \frac{\sin((2m+1)\pi u)}{2m+1} e^{-(2m+1)^2\pi^2 t}.
\end{equation}
For large $N$, the dominant contribution to the integral over $t$ stems from the small $t$ regime. This permits the replacement of the discrete summation with a continuous integral. Defining $k = (2m+1)\pi \sqrt{t}$, the discrete step is $\Delta k = 2\pi\sqrt{t}$. The sum becomes
\begin{equation}
\text{Inner sum} \simeq \frac{4}{\pi} \int_0^\infty \frac{dk}{2\pi\sqrt{t}} \frac{\sin(k \frac{u}{\sqrt{t}})}{k / (\pi\sqrt{t})} e^{-k^2} = \frac{2}{\pi} \int_0^\infty \frac{dk}{k} \sin\left(k \frac{u}{\sqrt{t}}\right) e^{-k^2}.
\end{equation}
Let us now define $z = u/\sqrt{t}$ and the integral above as $Y(z)$
\begin{equation}
Y(z) = \frac{2}{\pi} \int_0^\infty dk \, \frac{\sin(kz)}{k} e^{-k^2}.
\end{equation}
Taking the derivative with respect to $z$ cancels the $k$ in the denominator:
\begin{equation}
Y'(z) = \frac{2}{\pi} \int_0^\infty dk \, \cos(kz) e^{-k^2}.
\end{equation}
Using the standard Gaussian integral identity $\int_0^\infty \cos(ax) e^{-x^2} dx = \frac{\sqrt{\pi}}{2} e^{-a^2/4}$, we evaluate $Y'(z)$ to be
\begin{equation}
Y'(z) = \frac{2}{\pi} \left( \frac{\sqrt{\pi}}{2} e^{-z^2/4} \right) = \frac{1}{\sqrt{\pi}} e^{-z^2/4}.
\end{equation}
We recover $Y(z)$ by integrating $Y'(z)$. We utilize the boundary condition $Y(0) = 0$, establishing validity for domains including the origin ($0 \le u \le 1/2$)
\begin{equation}
Y(z) = \int_0^z \frac{1}{\sqrt{\pi}} e^{-z'^2/4} dz'.
\end{equation}
Applying the substitution $x = z'/2$ where $dz' = 2dx$, this transforms into the standard error function
\begin{equation}
Y(z) = \frac{2}{\sqrt{\pi}} \int_0^{z/2} e^{-x^2} dx = \text{erf}\left(\frac{z}{2}\right) = \text{erf}\left(\frac{u}{2\sqrt{t}}\right)
\end{equation}
Substituting $Y(z)$ back into the original expression for $I_N(u)$ yields:
\begin{equation}
I_N(u) \simeq \int_0^\infty dt \left[ \text{erf}\left(\frac{u}{2\sqrt{t}}\right) \right]^N \label{eq:7}
\end{equation}

\begin{figure*}
    \includegraphics[width=16cm]{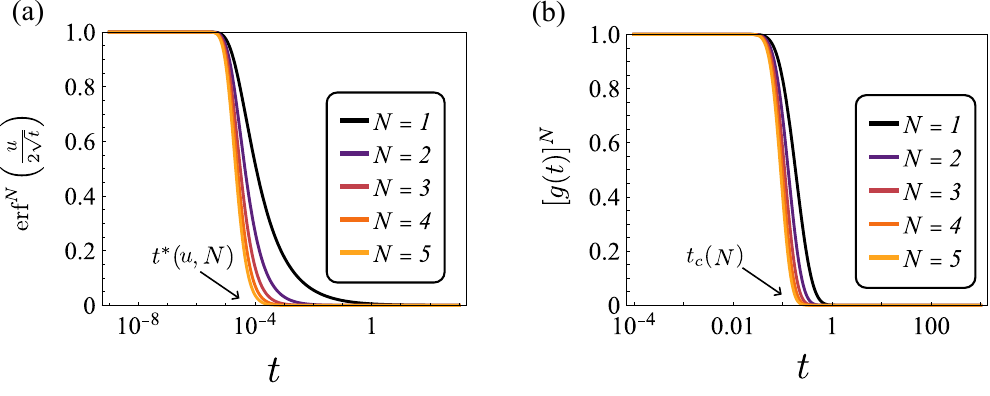}
    \caption{(a) Variation of $\text{erf}^N\left(\frac{u}{2\sqrt{t}}\right)$ with $u=10^{-2}$, and of (b) of $[g(t)]^N$ as a funtion of time $t$. As $N$ becomes larger, both the functions approach a step function with cut-offs $t^*(u,N)$ and $t_c(u,N)$, respectively.  }
    \label{step_fn}
\end{figure*}
As $N$ increases, the function $\left[ \text{erf}\left(\frac{u}{2\sqrt{t}}\right) \right]^N$ approaches a step function $\Theta[t^*(u,N) - t]$ (see \fref{step_fn}(a))
\begin{equation}
\text{erf}^N\left(\frac{u}{2\sqrt{t}}\right) \to \Theta[t^*(u,N) - t], \label{eq:8}
\end{equation}
where $t^*(u,N)$ can be estimated as follows. For small $t$ (important contribution comes from small $t$ as $N \to \infty$)
\begin{equation}
\text{erf}\left(\frac{u}{2\sqrt{t}}\right) \simeq 1 - \frac{e^{-\frac{u^2}{4t}}}{\sqrt{\pi} \frac{u}{2\sqrt{t}}}. \label{eq:9}
\end{equation}
Which implies
\begin{align}
\left[ \text{erf}\left(\frac{u}{2\sqrt{t}}\right) \right]^N &\simeq e^{-N e^{-\frac{u^2}{4t}}}, \nonumber\\
& \simeq e^{- e^{\ln N-\frac{u^2}{4t}}}.
\label{eq:10}
\end{align}
Evidently, to have a finite value of $\mathcal{O}(1)$ below $t<t^*$ we should have  $\frac{u^2}{4t^*} \simeq \ln N$, which gives $t^*(u,N) \simeq \frac{u^2}{4\ln N}$.
Now, in \eqref{eq:7}, we can cut the integrals off at $t^*(u,N)$ and obtain
\begin{equation}
I_N(u) \simeq \int_0^{t^*(u,N)} dt = t^*(u,N) \simeq \frac{u^2}{4\ln N}, ~~~ \text{for}~~~ 0 \le u\le 1/2.\label{eq:11}
\end{equation}
Finally, noting the symmetric property of $I_N$ we find
\begin{equation}
I_N(u) \simeq  \frac{(1-u)^2}{4\ln N}, ~~~ \text{for}~~~ 1/2 \le u\le 1.
\end{equation}
Combinedly, we have
\begin{equation}
I_N(u) \simeq  \frac{\text{min}(u,1-u)^2}{4\ln N},
\end{equation}
which holds for any arbitrary $0 \le u \le 1$.

\subsubsection{Derivation of $J_N(u)$ for $u \to 1$}
For this purpose, we set $u=1-\epsilon$ and later set $\epsilon \to 0$. From Eq. \eqref{eq:2} we get
\begin{equation}
J_N(1-\epsilon) = \frac{N 2^N}{\pi^{N-2}} \int_0^\infty dt \left[ \sum_{n=1}^\infty (-1)^{n-1} n \sin(n\pi\epsilon) e^{-n^2\pi^2 t} \right] \times \left[  \sum_{m=0}^\infty \frac{2\sin((2m+1)\pi\epsilon)}{2m+1} e^{-(2m+1)^2\pi^2 t} \right]^{N-1}
\end{equation}
But unlike in $I_N(1-\epsilon)$, we can now take directly the $\epsilon \to 0$ limit as the resulting integral over $t$ remains convergent (as will be seen shortly). Hence replacing $\sin(n\pi\epsilon) \simeq n\pi\epsilon$ and $\sin((2m+1)\pi\epsilon) \simeq (2m+1)\pi\epsilon$ to leading order in $\epsilon$, we arrive at
\begin{equation}
J_N(1-\epsilon) \simeq N 2^{2N-1} \epsilon^N \pi^2 \int_0^\infty dt \left[ \sum_{n=1}^\infty (-1)^{n-1} n^2 e^{-n^2\pi^2 t} \right] \times \left[ \sum_{m=0}^\infty e^{-(2m+1)^2\pi^2 t} \right]^{N-1} .\label{eq:12A}
\end{equation}
Let us define the first summation under the integral as $f(t)$ given below
\begin{equation}
f(t) = \sum_{n=1}^\infty (-1)^{n-1} n^2 e^{-n^2\pi^2 t}.
\end{equation}
As $t \to \infty$, $f(t) \simeq e^{-\pi^2 t}$. 
On the other hand, to deduce the small $t$ behavior of $f(t)$, we can use an alternative expression of $f(t)$ using the Jacobi identity
\begin{equation}
1 + 2\sum_{k=1}^\infty (-1)^k e^{-k^2 z} = 2\sqrt{\frac{\pi}{z}} \sum_{k=0}^\infty e^{-(2k+1)^2 \frac{\pi^2}{4z}}.
\end{equation}
Differentiating once with respect to $z$, one gets
\begin{equation}
\sum_{k=1}^\infty (-1)^{k-1} k^2 e^{-k^2 z} = -\frac{\sqrt{\pi}}{2z^{3/2}} \sum_{k=0}^\infty e^{-(2k+1)^2 \frac{\pi^2}{4z}} + \frac{\pi^{5/2}}{4z^{5/2}} \sum_{k=0}^\infty (2k+1)^2 e^{-(2k+1)^2 \frac{\pi^2}{4z}}.
\end{equation}
Choosing $z = \pi^2 t$ we get
\begin{equation}
f(t) = -\frac{1}{2\pi^{5/2} t^{3/2}} \sum_{k=0}^\infty e^{-\frac{(2k+1)^2}{4t}} + \frac{1}{4\pi^{5/2} t^{5/2}} \sum_{k=0}^\infty (2k+1)^2 e^{-\frac{(2k+1)^2}{4t}}. \label{eq:17}
\end{equation}
Note that $f(t)$ in \eqref{eq:17} is exact for all $t$. However, this representation is more suitable to extract the small $t$ behaviour, as only the $k=0$ term in the second sum is the dominant contribution. Hence, we get
\begin{equation}
f(t) \xrightarrow{t \to 0} \frac{1}{4\pi^{5/2} t^{5/2}} e^{-\frac{1}{4t}} \label{eq:18}.
\end{equation}
We now consider the second sum in \eqref{eq:12A}, namely the function
\begin{equation}
g_1(t) = \sum_{m=0}^\infty e^{-(2m+1)^2\pi^2 t}. \label{eq:20}
\end{equation}
As $t \to \infty$, $g_1(t) \simeq e^{-\pi^2 t}$. On the other hand, as $t \to 0$, one can again replace the sum by an integral after defining $(2m+1)\pi\sqrt{t} = y$, with $\Delta y = 2\pi\sqrt{t}$ being small for $t \to 0$. Hence
\begin{equation}
g_1(t) \xrightarrow{t \to 0} \frac{1}{2\pi\sqrt{t}} \int_0^\infty dy \, e^{-y^2} = \frac{1}{2\pi\sqrt{t}} \frac{\sqrt{\pi}}{2} = \frac{1}{4\sqrt{\pi t}} .\label{eq:22}
\end{equation}
Let us further define
\begin{equation}
g(t) = 4\sqrt{\pi t} g_1(t) = 4\sqrt{\pi t} \sum_{k=0}^\infty e^{-(2k+1)^2\pi^2 t}. \label{eq:23}
\end{equation}
such that $g(t) \to 1$ as $t \to 0$. The asymptotics of $g(t)$ are given by $g(t \to 0) \simeq 1 - 2 e^{-\frac{1}{4t}}$ (the second term can be obtained by using the Jacobi identity) and $g(t \to \infty) \to 4 \sqrt{\pi t}e^{-\pi^2 t}$.
Let us now express the integrand $J_N(1-\epsilon)$ in \eqref{eq:12A} in terms of $f(t)$ and $g(t)$. Following the same we get
\begin{equation}
J_N(1-\epsilon) \simeq N 2^{2N-1} \epsilon^N \pi^2 \int_0^\infty dt f(t) \left[ \frac{g(t)}{4\sqrt{\pi t}} \right]^{N-1} = \frac{2N \epsilon^N}{\pi^{(N-5)/2}} \int_0^\infty \frac{dt}{t^{(N-1)/2}} f(t) [g(t)]^{N-1}. \label{eq:25}
\end{equation}
Since $f(t) \sim \frac{e^{-\frac{1}{4t}}}{t^{5/2}}$ as $t \to 0$, it follows that the integrand is integrable at $t \to 0$ and finally it is integrable as $t \to \infty$. Hence, taking the $\epsilon \to 0$ limit is justified as argued at the beginning of this section.
We now analyze \eqref{eq:25} in the large $N$ limit. Consider the function $[g(t)]^{N-1} \approx [g(t)]^N$ with increasing $N$. It approaches a step function $\Theta[t_c(N) - t]$ (see \fref{step_fn}(b)). To estimate $t_c(N)$, we note that for large $N$, most important contribution to \eqref{eq:25} comes from the small $t$ regime.
For small $t$ we have
\begin{equation}
[g(t)]^N \simeq \left[ 1 - 2e^{-\frac{1}{4t}} \right]^N \simeq e^{-2N e^{-\frac{1}{4t}}}.
\end{equation}
The above expression approaches $1$ for $t < t_c(N) = \frac{1}{4\ln(2N)} \simeq \frac{1}{4\ln N}$, and approaches to $0$ for $t > t_c(N) \simeq \frac{1}{4\ln N}$. Hence, from \eqref{eq:25} we obtain
\begin{equation}
J_N(1-\epsilon) \simeq \frac{2N \epsilon^N}{\pi^{(N-5)/2}} \int_0^{t_c(N)} \frac{dt}{t^{(N-1)/2}} f(t). \label{eq:26}
\end{equation}
Using the small $t$ asymptotics of $f(t)$ from \eqref{eq:18}, we get
\begin{equation}
J_N(1-\epsilon) \simeq \frac{2N \epsilon^N}{\pi^{(N-5)/2}} \int_0^{\frac{1}{4\ln N}} \frac{dt}{t^{(N-1)/2}} \frac{1}{4\pi^{5/2} t^{5/2}} e^{-\frac{1}{4t}} = \frac{N \epsilon^N}{2\pi^{N/2}} \int_0^{\frac{1}{4\ln N}} \frac{dt}{t^{(N+4)/2}} e^{-\frac{1}{4t}} .\label{eq:27}
\end{equation}
Making a change of variable $z = \frac{1}{4t}$, $dt = -\frac{dz}{4z^2}$ we have
\begin{equation}
J_N(1-\epsilon) = \frac{N \epsilon^N}{2\pi^{N/2}} \int_{\ln N}^\infty \frac{dz}{4z^2} (4z)^{(N+4)/2} e^{-z} \simeq \frac{N 2^{N+1} \epsilon^N}{\pi^{N/2}} \int_{\ln N}^\infty dz \, z^{N/2} e^{-z}.
\end{equation}
The integral in the above expression can be written as 
\begin{align}
    \int_{\ln N}^\infty dz \,  e^{-z +\frac{N}{2}\ln z}=  \int_{\ln N}^\infty dz \,  e^{-\Phi(z,N)},
\end{align}
where $\Phi(z,N)=z-\frac{N}{2}\ln z$. To find the local minimum of the function $\Phi(z)$, we set $\Phi'(z)=0$ which occurs at $z=\frac{N}{2}$. We also have $\Phi''(z)=\frac{2}{N}>0$ which confirms that the extrema at $z=\frac{N}{2}$ is a local minima. Since, $N/2  >\ln N$ in the limit of $N \to \infty$, using saddle point approximation in the integral we eventually have
\begin{align}
    \int_{\ln N}^\infty dz \,  e^{-z +\frac{N}{2}\ln z}\simeq\sqrt{\pi N}\left(\frac{N}{2}\right)^{N/2}e^{- N/2}.
\end{align}
Plugging this result back to the expression for $J_N(1-\epsilon)$ we finally obtain
\begin{equation}
J_N(1-\epsilon) \simeq 2\sqrt{\pi}N^{3/2} \left(\frac{2\epsilon^2N}{\pi e}\right)^{N/2}. \label{eq:28}
\end{equation}

\subsubsection{Derivation of $J_N(u)$ for $u \to 0$}
To evaluate the limit of small $u \to 0$, we can independently approximate the two infinite sums within the integral brackets. Let us define the first sum as $h_1(u, t)$ and the second sum as $h_2(u, t)$. Let us first evaluate $h_1(u,t)$

\begin{equation}
h_1(u, t) = 2\pi \sum_{n=1}^\infty n \sin(n\pi u) e^{-n^2\pi^2 t}.
\end{equation}
For small $u$, we define the continuous scaling variable $y = n\pi u$. As $n$ changes by 1, $\Delta y = \pi u$. Because $u$ is small, $\Delta y$ is also small, which allows us to replace the summation with an integral

\begin{align}
h_1(u, t) &\simeq 2\pi \int_0^\infty \frac{dy}{\pi u} \left( \frac{y}{\pi u} \right) \sin(y) e^{- \frac{t y^2}{u^2}}, \\
&\simeq \frac{2}{\pi u^2} \int_0^\infty dy \, y \sin(y) e^{- \frac{t y^2}{u^2}}.
\end{align}
Using the exact Gaussian integral result $\int_0^\infty y \sin(y) e^{-a y^2} dy = \frac{\sqrt{\pi}}{4 a^{3/2}} e^{-1 / 4a}$, we can set $a = \frac{t}{u^2}$ to have
\begin{equation}
h_1(u, t) \simeq \frac{2}{\pi u^2} \left( \frac{\sqrt{\pi} u^3}{4 t^{3/2}} e^{-\frac{u^2}{4t}} \right) = \frac{u}{2\sqrt{\pi} t^{3/2}} e^{-\frac{u^2}{4t}}.
\end{equation}
The second term $h_2(u,t)$ could be found in a similar way
\begin{equation}
h_2(u, t) = \frac{2}{\pi} \sum_{n=1}^\infty \frac{1-(-1)^n}{n} \sin(n\pi u) e^{-n^2\pi^2 t}\simeq \text{erf}\left(\frac{u}{2\sqrt{t}}\right).
\end{equation}
We now substitute the asymptotic forms of $h_1$ and $h_2$ back into the original equation for $J_N(u)$ as
\begin{equation}
J_N(u) \simeq N \int_0^\infty dt \left( \frac{u}{2\sqrt{\pi} t^{3/2}} e^{-\frac{u^2}{4t}} \right) \left[ \text{erf}\left(\frac{u}{2\sqrt{t}}\right) \right]^{N-1}.
\end{equation}
To solve this, we again perform another change of variables by setting $z = \frac{u}{2\sqrt{t}}$. This substitution implies that $t = \frac{u^2}{4z^2}$, and differentiation gives $dt = -\frac{u^2}{2z^3} dz$. The integration bounds map from $t \in [0, \infty)$ to $z \in [\infty, 0]$ so that
\begin{equation}
J_N(u) \simeq N \int_\infty^0 \left( -\frac{u^2}{2z^3} dz \right) \left( \frac{u}{2\sqrt{\pi} \left( \frac{u^2}{4z^2} \right)^{3/2}} e^{-z^2} \right) [\text{erf}(z)]^{N-1}.
\end{equation}
Simplifying the constants and the integrand gives
\begin{equation}
J_N(u) \simeq N \int_0^\infty dz \left( \frac{2}{\sqrt{\pi}} e^{-z^2} \right) [\text{erf}(z)]^{N-1}.
\end{equation}
Recognizing that the standard derivative of the error function is $\frac{d}{dz} \text{erf}(z) = \frac{2}{\sqrt{\pi}} e^{-z^2}$, the integral simplifies perfectly to
\begin{align}
J_N(u) &\simeq \int_0^\infty N [\text{erf}(z)]^{N-1} d(\text{erf}(z)) \\
&\simeq [\text{erf}(z)]^N \Big|_0^\infty.
\end{align}
Because $\text{erf}(\infty) = 1$ and $\text{erf}(0) = 0$, evaluating the bounds yields the final result
\begin{equation}
J_N(u \to 0) \approx 1.
\end{equation}

\newpage
\bibliographystyle{unsrt}
\bibliography{fpusr-1}

\end{document}